\documentclass[usenatbib,useAMS]{mnras}

\usepackage[T1]{fontenc}
\usepackage{ae,aecompl}

\usepackage{graphicx}	
\usepackage{amsmath}	
\usepackage{amssymb}	

\usepackage{epsfig}
\usepackage{longtable}
\usepackage[usenames, dvipsnames]{color}

\usepackage{natbib}
\usepackage{pifont}
\usepackage{txfonts}
\usepackage{mathrsfs}

\title[Cloud formation in metal-rich planets]{Cloud formation in metal-rich atmospheres\newline of hot super-Earths like 55 Cnc e and CoRot7b}

\author[G.Mahapatra et al.]{
G.Mahapatra,$^{1}$
Ch.Helling,$^{1}$\thanks{E-mail: ch80@st-andrews.ac.uk}
and Y. Miguel$^{2}$
\\
$^{1}$Centre for Exoplanet Science, SUPA, School of Physics and Astronomy, University of St. Andrews, St. Andrews, UK, KY16 9SS\\
$^{2}$Laboratoire Lagrange, UMR 7293, Universite de Nice-Sophia Antipolis, CNRS, Observatoire de la C\^ote d'Azur,\\ Blvd de l'Observatoire, CS 34229, 06304 Nice cedex 4, France
}

\date{Accepted XXX. Received YYY; in original form ZZZ}

\pubyear{2016}

\begin{document}
\maketitle
\sloppy

\begin{abstract}

Clouds form in the atmospheres of planets where they can determine the observable spectra, the albedo and phase curves.  Cloud properties are determined by the local thermodynamical and chemical conditions of an atmospheric gas. A 
retrieval of gas abundances requires a comprehension of the cloud
formation mechanisms under varying chemical conditions.  With the aim of studying cloud formation in metal rich atmospheres, we explore the possibility of clouds in evaporating exoplanets like
CoRoT-7b and 55 Cnc e in comparison to a generic set of solar
abundances and the metal-rich gas giant HD149026b.  We assess the impact
of metal-rich, non-solar element abundances on the gas-phase chemistry, and  apply our
kinetic, non-equilibrium cloud formation model to study
cloud structures and their details. We provide an overview of global cloud properties in terms of material compositions, maximum particle formation rates, and average cloud particle sizes for various sets of rocky element abundances.
Our results suggest that the conditions on 55 Cnc e
and HD149026b should allow the formation of  mineral clouds in their atmosphere.  The high temperatures on  some hot-rocky super-Earths (e.g. the day-side of Corot-7b) result in an ionised atmospheric gas and they  prevent gas condensation, making cloud formation unlikely on its day-side. 

\end{abstract}




\section{Introduction}


 With telescopes of high sensitivity and sophistication in place, we
 have ventured into an era of observing and characterizing the
 atmospheres of exoplanets in greater detail. Based on mass and
   radius measurements, bulk densities are used to draw first
   conclusions about possible planetary chemical compositions: A rocky
   bulk composition of an Earth-like silicate/iron mixture appears
   consistent with masses and radii of short period planets if their
   radius $<1.5$R$_{\rm Earth}$ (\citealt{2015ApJ...800..135D}). For
   radii $>1.5$R$_{\rm Earth}$, bulk densities appear so low that most of these
   planets must have a large volatile atmosphere
   (\citealt{2015ApJ...801...41R}). The observation and
 characterization of these atmospheres could provide a unique
 opportunity to study the internal composition and surface material of
 rocky exoplanets (\citealt{miguel2011}) which is largely
   unknown.  Planetary atmospheres with local gas temperature $>$1000
 K have a rich molecular chemistry and they form silicate clouds
 similar to brown dwarfs
 (e.g. \citealt{sudarsky2005,fortney2008,witte2009dust}).  Clouds alter the observed spectra by flattening the
 ultra-violet (UV) and visible spectrum due to scattering from small
 sized particles, having a cooling effect on the atmosphere beneath
 the optically thick region by reflecting the received spectra and
 also by depleting the local gas-phase of minerals to condense and
 form mineral cloud particles \citep{helling2014atmospheres}.

Even though it is crucial to provide a proper interpretation of
observations, cloud formation processes in hot and warm super-Earth
atmospheres are poorly studied. Clouds have been proposed for
super-Earth GJ436\,b wherein the observed spectrum in
1.2$\,\ldots\,$1.6 $\mu$m is featureless indicating a possibility of a
"heavier than hydrogen" gas-phase molecular composition resulting in
high opacity clouds that form at a pressure altitude of
$\sim$10$^{-3}$ bars
\citep{knutson2014featureless}. \cite{fegley2016solubility} study the
surface out-gassing of hot rocky exoplanets taking into account the
fractional vaporization assuming the gas contains the monomers that
the rocks are made of. 
\cite{schaefer2012vaporization} and \cite{schaefer2009}
discuss the possible atmosphere of Corot-7b like planets. They suggest
the presence of Na and K clouds in the upper parts of the atmosphere.
\cite{miguel2011} explore the composition of initial planetary
atmospheres of hot rocky super-Earths (HRSE) in the Kepler planet candidate
sample. \cite{Itoetal} conducted a similar study with a few more types
of Earth rock compositions. \cite{demory2016map} have derived a
longitudinal temperature map of the super-Earth 55 Cancri-e and find a
large temperature contrast of 1300 K between its day and night
sides. Although they suggest the temperature map to be a surface map
of the planet, \cite{demory2016map} do not rule out the possibility of
a dense atmosphere with thick clouds on the day side.

The aim of this paper is to explore cloud formation in hot and highly
enriched exoplanet atmospheres. Atmospheres of small hot planets
  are poorly understood so that little is known about their composition,
  chemistry and condensibles. We therefore explore a diversity of
  scenarios, including extremes. Beside addressing specific
  objects (CoRoT-7b, 55 Cnc e), this strategy allows to understand our
  results in a large context by comparing to known and better studied
  cases (like brown dwarfs which exhibit comparable thermodynamic
  properties to giant gas planets), and to identify trends regarding cloud properties and
  the effect of element abundances on cloud properties. We note for
  clarification that, for example, hydrogen-rich atmospheres for
  planets like Corot-7b and 55 Cnc e are unlikely unless they are very
  young as studied by \cite{2016MNRAS.459.4088O}.
  \cite{2016arXiv161009390L}'s Fig.~1 suggest for a planet of
  1.58R$_{\rm Earth}$ (CoRoT-7b's radius) that a percentage of H/He of
  less than 0.1\% remains in the atmosphere after 5Gyr for all their
  highly irradiated cases.

 Here we suggest that a rich, non-homogeneous cloud chemistry is
 to be expected for non-solar and for processed atmospheric element
 abundances as originating from evaporating planet surfaces or as a
 result of disk evolution
 (e.g. \citealt{hell2014,modisini2016,eistrup2016,crid2016}) that
 might lead to locally high metallicities in the gas being accreted by
 the planets. We  explore the evaporating planet CoRoT-7b and the
 magma-planet 55 Cnc e for various Earth rock compositions as examples
 for possibly processed, non-solar element abundances. We compare
 these results with HD149\,026b for which a $10$ times solar
 metallicity in the element abundances was tested
 (\citealt{fort2006}), and use a standard solar element abundance  for non-irradiated
 giant-gas planet atmosphere as reference. We do not consider the
   physical processes that lead to planetary mass loss which we assume
   is launched above our computational domain. Intense X-ray and EUV
   irradiation (driving intense heating and ionisation) in combination with centrifugal acceleration can lead to a hydrodynamic mass loss from the upper atmospheres of hot
   Jupiters    (\citealt{2015ApJ...813...50K}). Tidal forces will decelerate a
   wind and MHD effects will confine the wind further to the volume of closed field lines.
   (\citealt{2011ApJ...728..152T}). Due to the complexity of the
   planetary mass loss models, no consistent link has been made to the
   underlying atmosphere structure where cloud formation affects the atmosphere structure and determines 
   the observable spectrum, and the inner boundary is not well defined
   in these complex models. Section \ref{sec:method} describes our
 approach. We summarise our kinetic cloud formation model and the input properties used. We note that our model neither prescribed the number of seed particles nor the particle size distributions (e.g. \citealt{hell2013}). In
 Sect.~\ref{sec:gasphase}, we investigate the gas phase composition
 for different element abundances as pre-requisite for the clouds to
 form.  Section \ref{sec:clouds} describes the cloud results obtained
 for different exoplanetary cases (CoRoT-7, 55 Cnc e, HD149\,026b,
 reference model) and the influence of different initial element
 abundances on the cloud structure. Section \ref{sec:summary} contains
 our conclusions.


\section{Approach}\label{sec:method}
We apply our kinetic, non-equilibrium cloud formation model to
investigate cloud structures in atmospheres of non-solar element
abundances.  We assess the impact of element abundances on the
gas-phase chemistry and the cloud structure details for prescribed
planetary atmosphere profiles. 

\subsection{Gas chemistry and kinetic cloud modelling}\label{ss:cloudinp}

\paragraph*{Gas-phase chemistry:}
 We apply a chemical equilibrium approach to study the initial,
 undepleted gas composition for given atmospheric structures (T$_{\rm
   gas}$, p$_{\rm gas}$) and the various sets of element abundances
 described below. We use the code described in \cite{bilger2013}. The
 cloud formation code contains a chemical equilibrium routine
 similar to \cite{bilger2013} but without the  large carbo-hydrate molecules.

\paragraph*{Cloud formation model:}
Our cloud formation model describes the formation of clouds by
nucleation, subsequent growth by chemical surface reactions ontop of
the seeds, evaporation, gravitational settling, element
depletion/enrichment and convective replenishment
(\citealt{woi2003,woi2004,hell2006,helling2008}). The effect of
nucleation, growth \& evaporation on the remaining elements in the gas
phase is fully accounted for (Eqs. 10 in \citealt{helling2008}). The C/O
ratio at the start of the cloud formation process is fixed at C/O=0.5
for all of the initial cloud formation cases so as to be able to
discern the differences in the chemistry due to varying silicate
compositions. The surface growth of a diversity of materials causes
the grains to grow to $\mu$m-sized particles of a mixed
composition. We consider the reactions forming 12 different dust
species (TiO$_2$[s], Al$_2$O$_3$ [s], CaTiO$_3$[s], Fe$_2$O$_3$[s],
FeS[s], FeO[s], Fe[s], SiO[s], SiO$_2$ [s], MgO[s], MgSiO$_3$[s],
Mg$_2$SiO$_4$ [s]) which react to form grain mantles of a mix of these
species depending on the local gas density and temperature. The {\sc
  Drift} model of \cite{hell2006} used 60 growth reactions. We treat a
total of 79 surface growth reaction in this paper which are provided
in Table~\ref{table:reactions}.  The seed formation is described by
classical nucleation theory which has been modified for TiO$_2$
nucleation to take into account knowledge about (TiO$_2$)$_{\rm N}$
cluster formation (\citealt{jeong1999,jeong2000,lee2015a}).  A
  more extensive disucssion of the theoretical background of seed
  formation is provided in Sect. 3a in \cite{hell2013}. We note that
  photoionisation may open reaction paths that enable the formation of
  complex molecular or cluster as precoursers for seed formation at even
  lower temperatures and densities than considered here
  (\citealt{2011Natur.476..429K, 2014IJAsB..13..173R}). Photochemistry
  is not part of our present cloud formation model.

\begin{table*}
\centering
\caption{Initial element abundances, $\epsilon^{0}_{i}$,  from magma compositions. Solar and meteorite abundances are shown as a comparison.} 
\hfill \break
\resizebox{\linewidth}{!}{%
\begin{tabular}{c c c c c c c c c}

    \hline \hline
    Element Abundance & Komatiite$^{a}$ 
    & BSE$^{b}$ & BSE$^{c}$ & MORB$^{c}$ 
    & Upper Crust$^{d}$  & Bulk Crust$^{d}$ & Solar$^{e}$ & Meteorite$^{e}$\\
    log($\epsilon_{i}/\epsilon_{H}$) & & & & & & & &\\
    \hline

    Si & 7.54 & 7.54 & 7.54 & 7.54 & 7.54 & 7.54 & 7.54 & 7.55\\
    O & 8.09 & 8.10 & 8.10 & 7.97 & 8.00 & 7.97 & 8.69 & 8.43\\
    Mg & 7.52 & 7.62 & 7.64 & 7.01 & 6.61 & 6.29 & 7.54 & 7.56\\
    Fe & 6.86 & 6.72 & 6.72 & 6.68 & 6.51 & 6.34 & 7.45 & 7.49\\
    Ca & 6.63 & 6.48 & 6.46 & 6.42 & 6.59 & 6.30 & 6.36 & 6.33\\
    Al & 6.54 & 6.62 & 6.61 & 6.57 & 7.03 & 6.97 & 6.47 & 6.46\\
    Na & 5.82 & 5.71 & 5.73 & 5.69 & 6.53 & 6.52 & 6.33 & 6.30\\
    Ti & 5.12 & 5.01 & 5.06 & 5.66 & 5.49 & 5.39 & 5.02 & 4.95\\
    K & 4.93 & 4.56 & 4.47 & 4.79 & 6.12 & 6.27 & 5.08 & 5.11\\
    Cr & - & - & 5.36 & 4.62 & - & - & 5.64 & 5.67\\
    P & - & - & 4.12 & 4.77  & 4.79 & 4.82 & 5.36 & 5.44\\
    \hline
\end{tabular}}
{\small
$^{a}$ \cite{2004Icar..169..216S},
$^{b}$ O'Neill $\&$ Palme (1998),
$^{c}$ \cite{McDonough1995223},
$^{d}$ \cite{2009pctc.book.....T}.
$^{e}$ \cite{grevesse2007solar}.}
\label{tab:EAb}
\end{table*}

\subsection{Input properties}

\subsubsection{Element abundances}\label{ss:elm}

The element abundances are input properties that determine the
gas-phase composition of the atmosphere and therefore also the cloud
particle material composition. In this study we consider the following
species to model exoplanet atmospheres: H, He, Li, C, N, O, Fl, Ne,
Na, Mg, Al, Si, S, Cl, K, Ca, Cr, Mn, Fe, Ni, Ti. The undepleted
element abundances are called initial abundances
$\epsilon^{0}_{i}$. The close-in planets have an extremely hot surface
that will melt forming a lava ocean that can vaporise, resulting in an
enrichment of the initial composition of the atmosphere. Since the
lava ocean composition for an exoplanet is poorly constrained, we
adopt different potential compositions for a planetary crust, that
will result in different elemental abundances for our enriched
atmosphere. The enriched elements are O, Mg, Al, Si, Ca, Ti, Fe, and
their initial abundances, $\epsilon^{0}_{i}$, are adopted according to
the different compositions for Earths planetary crust tested in the
paper: bulk crust, upper crust, MORB - metal-oxide-ridge basalt, BSE -
bulk silicate Earth, komatite (see Table \ref{tab:EAb} and Sect.~\ref{ss:appA}).  All other
elements are set to solar values initially, except if all elements are
increased by the same factor. Our test cases remain hydrogen rich,
unless we specifically test a decreased hydrogen content as in the
case 'BSE (low H)'  (10$^{-5}$n$_{\rm H}$). The cloud formation
process changes the initial element abundances by nucleation, growth
and evaporation with respect to those elements that are involved (O,
Mg, Al, Si, S, Ca, Ti, Fe; see Eqs. 4 $\&$ 8 in \cite{woi2004}).

 Different sets of initial element abundances should influence the
  local temperature and pressure structure due to changing gas and
  cloud opacities.  Differences in element abundances will also affect
  the mean molecular weight. 

\begin{figure}
\hspace*{-0.4cm}
\centering
\includegraphics[scale = 0.37]{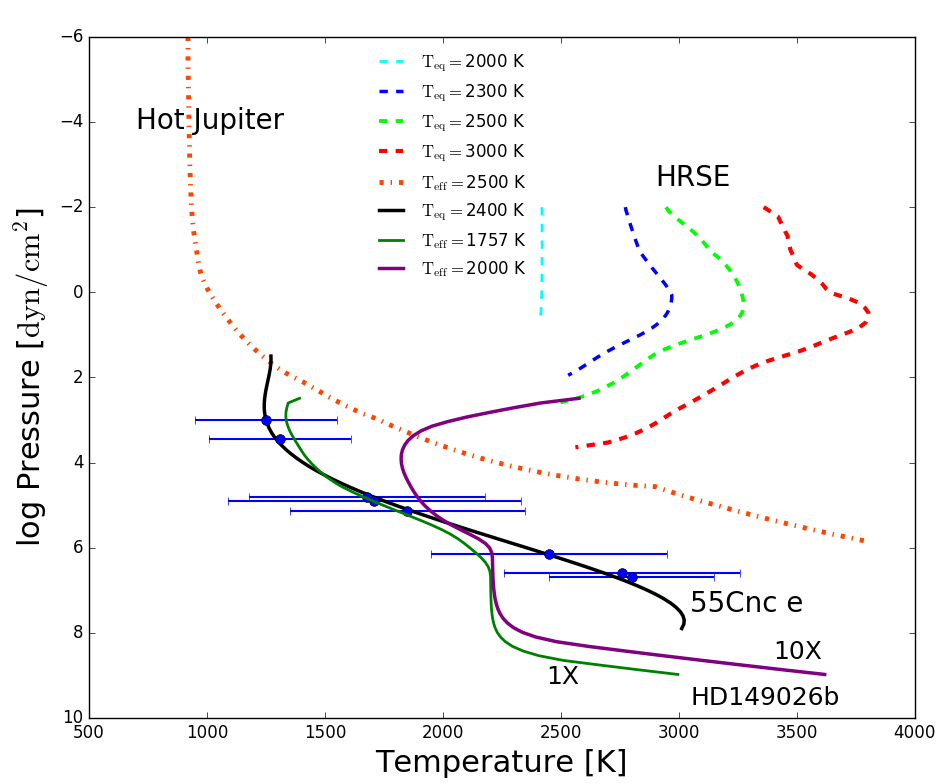}
\caption{The (T$_{\rm gas}$, p$_{\rm gas}$)-profiles of different
  planetary atmospheres. \newline {\sc Drift-Phoenix} models with
  T$_{\rm eff}$=2500K, log(g)=3.0 (orange dashed line;
  \citealt{witte2009dust,witte2011dust}). Hot rocky super-earths
  (HRSE) atmosphere profiles are demonstrated for T$_{\rm eq}$= 2000
  to 3000 K (dashed light blue, blue, green and red lines;
  \citealt{Itoetal}). HD149026b with $1\times$ and $10\times$ solar
  element abundances (solid purple and green;
  \citealt{fortney2006atmosphere}). 55 Cnc e (solid black and blue
  points with error bars, \citealt{demory2016variability}).}
\label{TPall}
\end{figure}

\subsubsection{Atmosphere profiles}\label{ss:atm}


We utilise  pre-calculated thermodynamic structure of atmosphere
  models. The resulting limitation of our approach is therefore that
  changes of initial element abundances (Sect.~\ref{ss:elm}) do not
  affect the (T$_{\rm gas}$, p$_{\rm gas}$) structure, nor will the
  forming cloud particles have an impact on the atmosphere structure
  considered. We note that the (non-irradiated) {\sc Drift-Phoenix}  giant gas planet
  model and the models for HD\,149026b, that we utilise in this paper,  do
  include the radiative and chemical feedback of clouds on the
  atmosphere structure. The profile used for 55 Cnc e results from a
  retrieval approach and should therefore also mimic the presence of
  clouds in case they are present. We use the following planetary
atmosphere profiles for our study of cloud formation in atmospheres
with metal enriched abundances: \\ -- non-irradiated giant gas
planet as reference and comparison to previous works\\ ({\sc
  Drift-Phoenix} model atmosphere: T$_{\rm eff}$=2500K,
log(g)=3.0),\\ -- the metal-rich mini-giant planet HD 149\,026b\\ (T$_{\rm
  eff}$=1757K, log(g)=3.23; \citealt{fort2006}), \\ -- CoRoT-7b
 as hot super-Earth example with T$_{\rm equ}$=2300K,
log(g)=3.62,  adopting the atmosphere model results from \cite{Itoetal} , and\\ -- 55 Cancri e profile as hot-rocky super-Earth
example adopting a retrieved  profile (T$_{\rm equ}$=2400K, log(g)=3.33;
\citealt{demory2016variability}).\\ Figure~\ref{TPall} compares these
different type of atmosphere model structures. Note that
  different communities use different global temperature definition\footnote{ The sub-stellar point equilibrium temperature is given as,
  $$ \mathrm{T_{eq}^4 = (1-A_P) \frac{R_{*}^{2}}{D^2} T_{*}^{4}} $$
  where $R_*$ and $T_*$ are respectively, the radius and temperature
  of the host star, $A_P$ is the planetary albedo, and D is the
  orbital distance of the planet. The host star is assumed to emit
  blackbody radiation of 6000 K and the magma composition is assumed
  to be BSE. }.  The 1D (T$_{\rm gas}$, p$_{\rm gas}$)
structures for 55Cnc e, HD 149\,026b and that for the hot non-irradiated giant gas
planet for T$_{\rm eff}$=2500K, log(g)=3.0 populate the same part of
the (T$_{\rm gas}$, p$_{\rm gas}$) diagram. The atmospheres profile
that was suggested for CoRoT-7b by \cite{Itoetal}  (T$_{\rm equ}$=2300K) is far hotter
than the gas giant atmosphere profiles. The T$_{\rm eff}=2500$K for
the giant-gas planet is chosen since it has a range of local gas
temperatures similar to that of hot super Earths or gas-giant planets
that we wish to study. CoRoT-7b and 55 Cnc e receive a comparable amount of irradiating flux as they orbit similar stars at a similar distance. The main differences between the two cases are therefore due to the different atmosphere profiles presently available in the literature and potential differences in their atmospheric compositions (55 Cnc e has a larger radius than Corot-7b consistent with a more extended atmosphere). The main differences between   We have run our cloud formation model for
  all these model atmosphere with all element abundances listed in
  Table~\ref{tab:EAb}, but we will later confine our results presentation
  to a selected set of input atmosphere structures.

\begin{table}
\caption{Global parameters of the planets considered in the paper.\label{planetschar}} 
\resizebox{0.5\textwidth}{!}{
\hspace*{-1cm}
\begin{tabular}{c c c c c c}
    \hline \hline
    Planets & a    & T$_{\rm global}$ & M            & log(g)     & References\\
            & [AU] & [K]         &  [M$\oplus$] & [cm/s$^2$] & \\
    \hline
    giant gas planet & - &  T$_{\rm eff}$=2500 & - & 3.0 & \cite{witte2009dust}\\ 
    55 Cnc e & $\sim$0.015 &  T$_{\rm equ}$=2400 & 8.63 & 3.33 & \cite{demory2016variability}\\
    HD149\,026b  & $\sim$0.042 &  T$_{\rm eff}\sim$1800 & 114 & 3.23 & \cite{fortney2006atmosphere}\\
    CoRoT-7b  & $\sim$0.017 &  T$_{\rm emu}$=2300 & $<$9 & $\sim$3.5 & \cite{Itoetal}\\

    \hline
    \end{tabular}}
\end{table}

\begin{figure*}
\centering
\includegraphics[scale = 0.5]{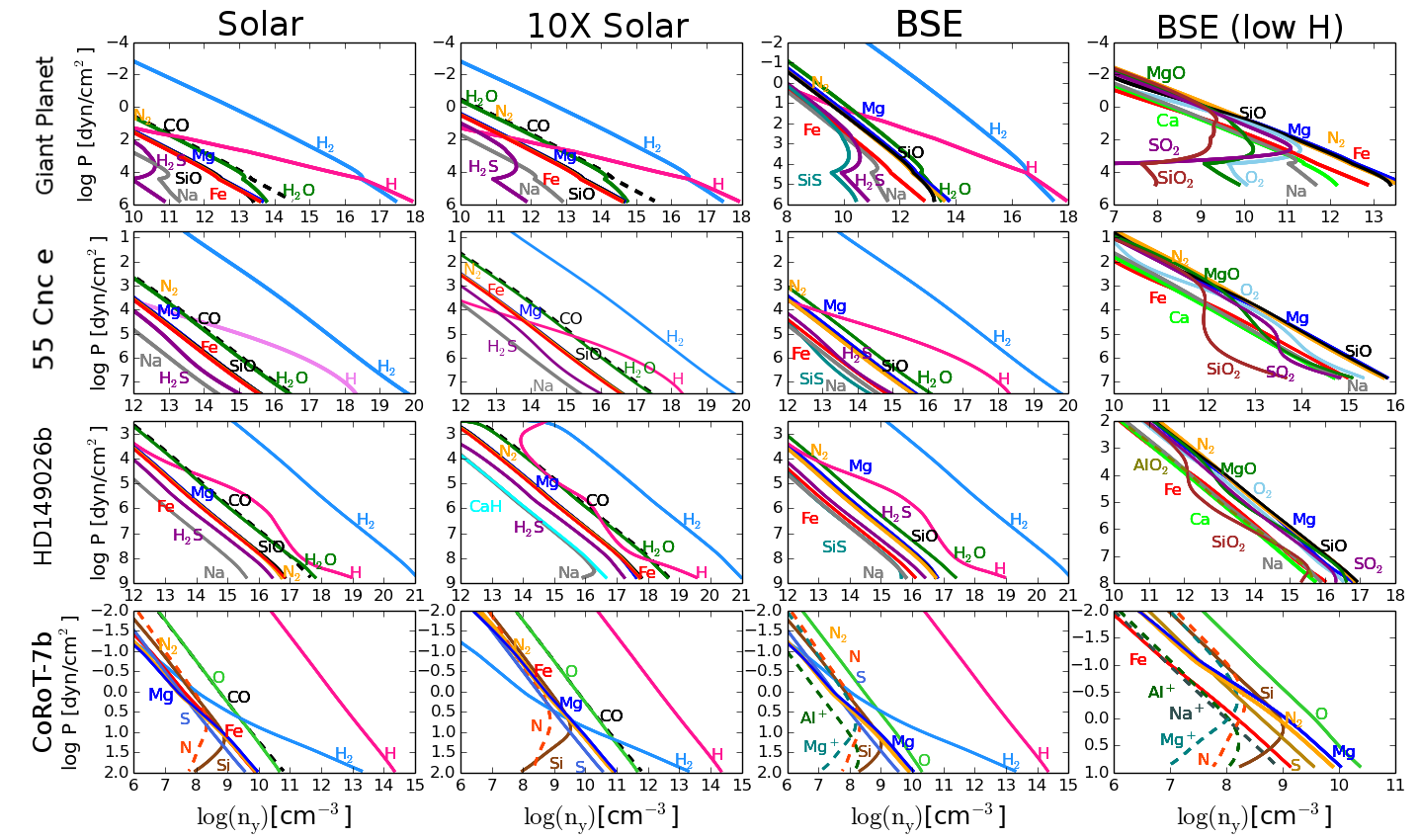}
\caption{The gas-phase number densities, n$_{\rm y}$  [cm$^{-3}$], of the 10 most abundant gas species  for four types of elements abundances  and four different (T$_{\rm gas}$, p$_{\rm gas}$)-profiles.  Different rows represent different atmospheric models as indicated and the respective   (T$_{\rm gas}$, p$_{\rm gas}$)-profiles are shown in Fig.~\ref{TPall}. The name of each gas species is indicated near the line with the same color.
Different columns represent different  sets of initial element abundances: solar, $10\times$solar -- $10\times \epsilon_{\rm solar}$,  BSE(low H) -- BSE abundance with $10^{-5}\times\epsilon_{\rm H}$. }
\label{Top10Gas}
\end{figure*}

\begin{figure*}
\centering
\includegraphics[scale = 0.4]{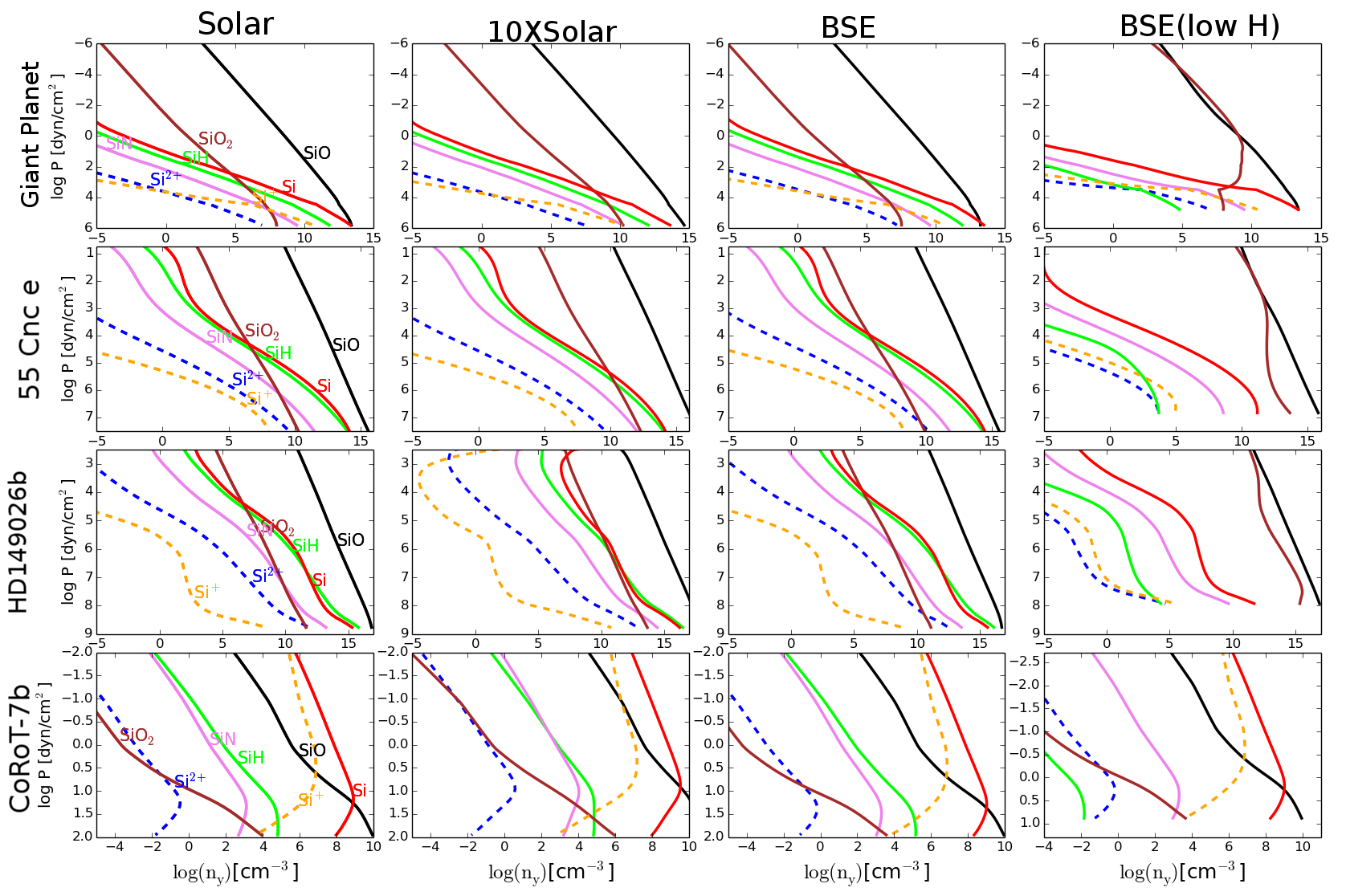}
\caption{The gas-phase number densities,  n$_y$ [cm$^{-3}$],  for Si-binding species in chemical  equilibrium  for four types of elements abundances  and four different (T$_{\rm gas}$, p$_{\rm gas}$)-profiles.  Different rows represent different atmospheric models as indicated and the respective   (T$_{\rm gas}$, p$_{\rm gas}$)-profiles are shown in Fig.~\ref{TPall}. Different columns represent different  sets of element abundances: $10\times$solar -- $10\times \epsilon_{\rm solar}$,  BSE(low H) -- BSE abundance with $10^{-5}\times\epsilon_{\rm H}$. All species have the same line color in all panels and are provided in column 1.}
\label{Si_All}
\end{figure*}

\begin{figure*}
\centering
\includegraphics[scale = 0.4]{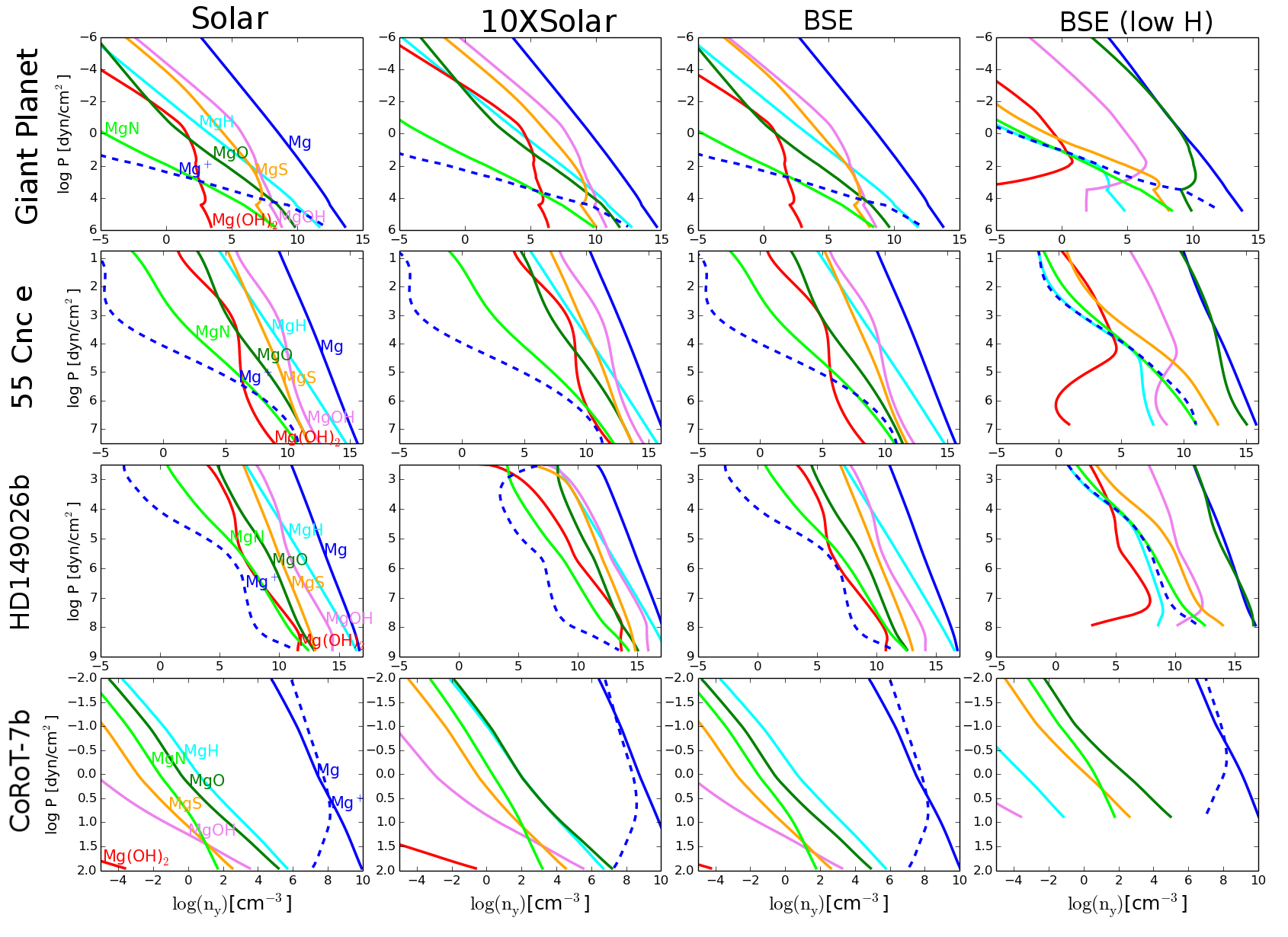}\\
\includegraphics[scale = 0.43]{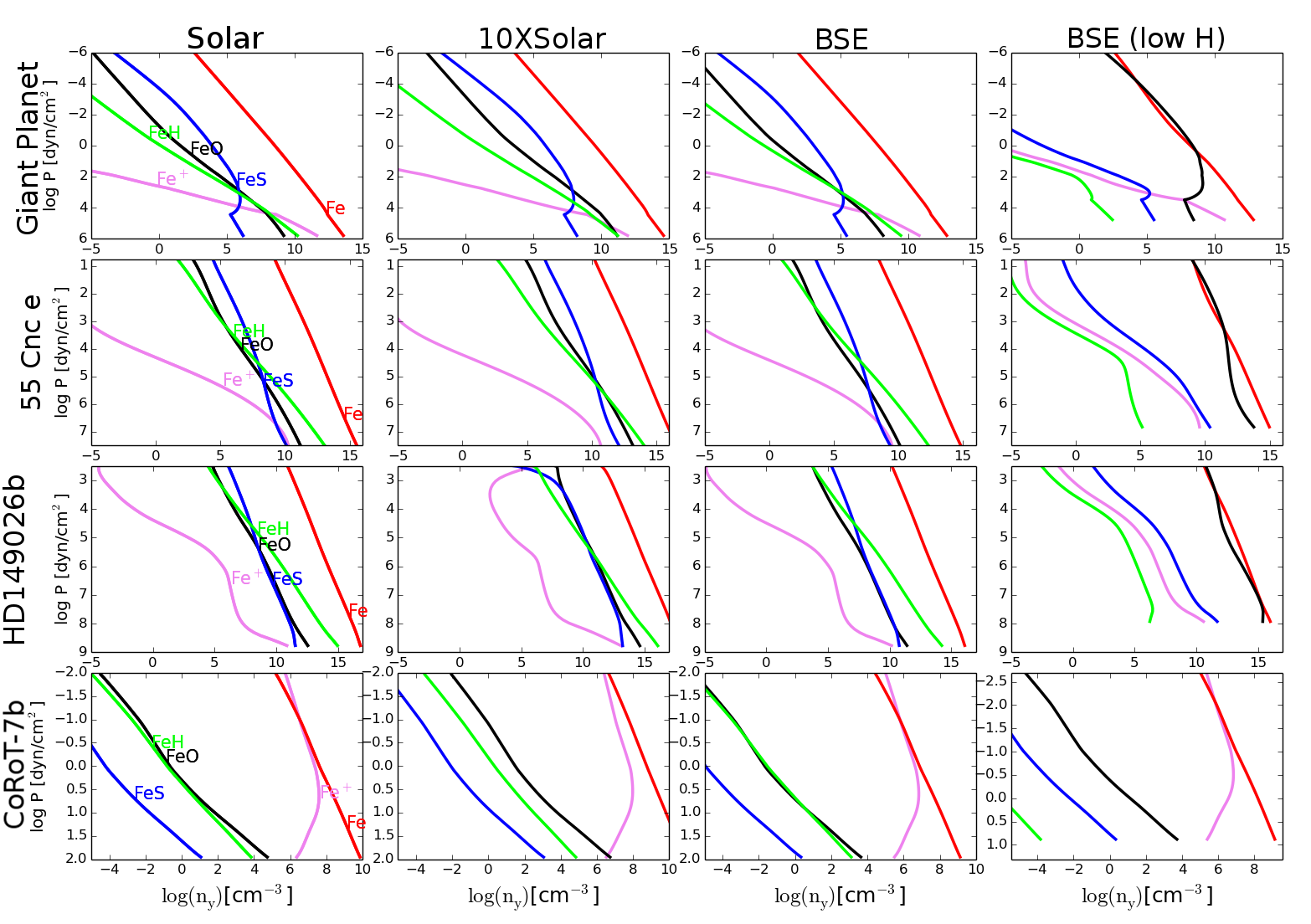}
\caption{Same like Fig.~\ref{Si_All} but for Mg-binding (top) and Fe-binding (bottom) gas-phase species.}
\label{Mg_All}
\end{figure*}

\begin{figure*}
\centering
\includegraphics[scale = 0.41]{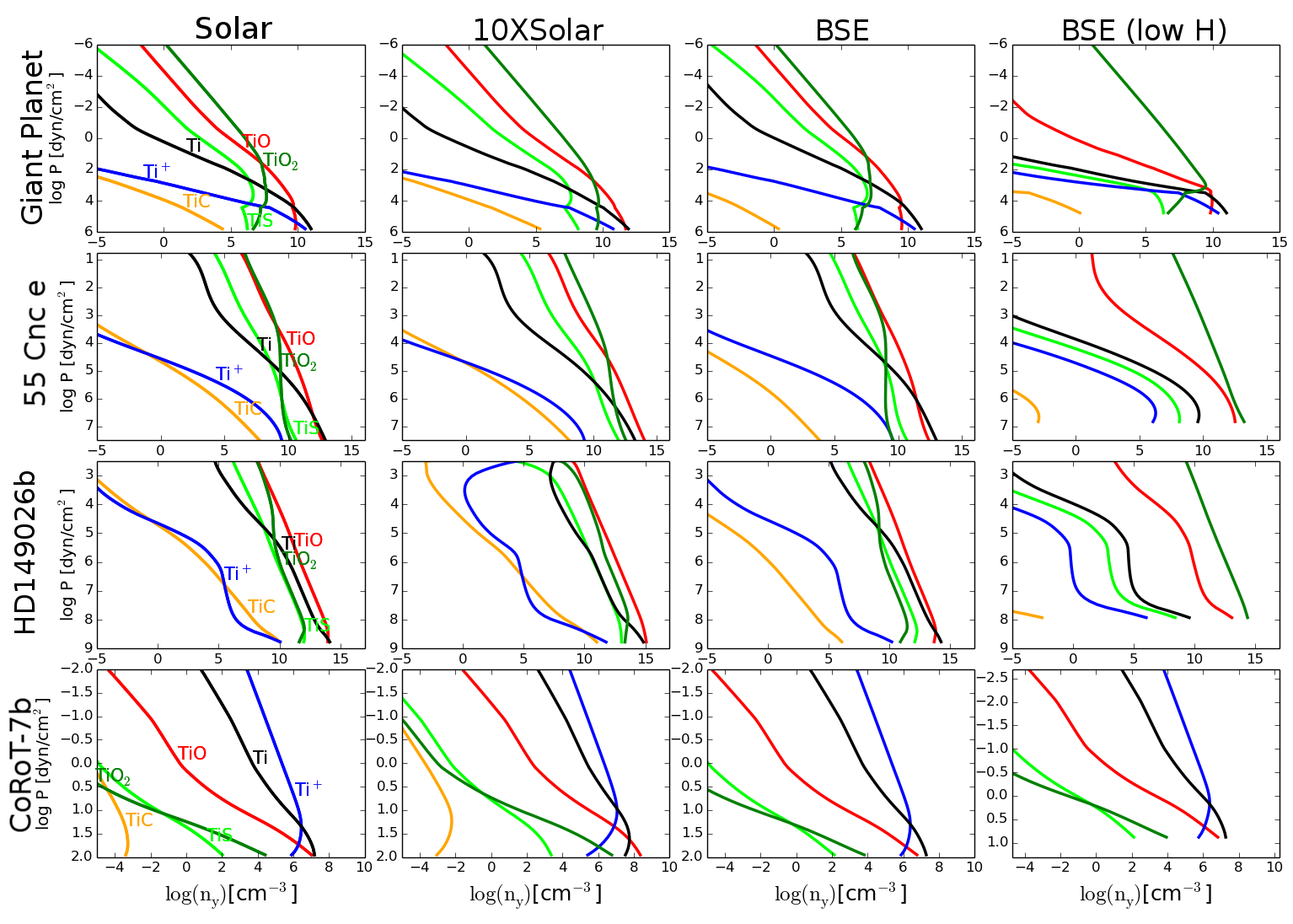}\\
\includegraphics[scale = 0.43]{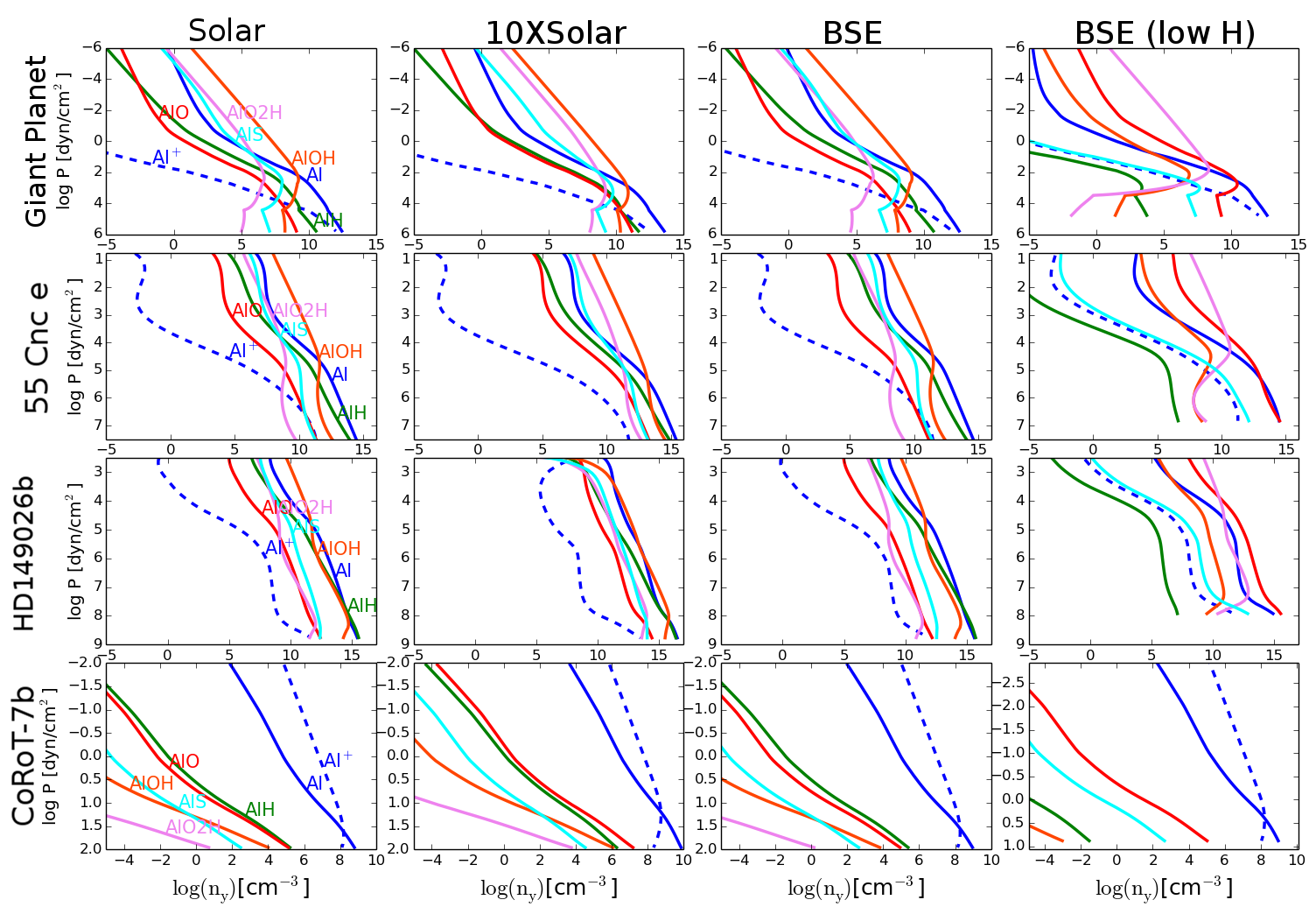}
\caption{Same like Fig.~\ref{Si_All} but for Ti-binding (top) and Al-binding (bottom)  gas-phase species.}
\label{Ti_All}
\end{figure*}

\paragraph*{Giant gas planet:}

We use a {\sc Drift-Phoenix} model atmosphere with a surface gravity,
log(g) = 3.0 and effective temperature, T$_{\rm eff}$ = 2500 K for our
cloud formation study. Figure \ref{TPall} shows the atmospheric
profiles of two different hot-Jupiters that are derived from the {\sc
  Drift-Phoenix} model atmosphere simulations described in
\cite{witte2009dust,witte2011dust} for comparison. The atmospheric
model used is in local thermal equilibrium (LTE), and the radiative
 and convective energy transport is solved to determine the local
gas temperature. The local gas pressure is calculated assuming
hydrostatic equilibrium, and the equations of state, opacity data and
equilibrium chemistry calculations close this system of equations; see
e.g.  \citep{helling2014atmospheres}.

\paragraph*{Metal-rich mini-giant planets, e.g.  HD\,149\,026b:}
HD 149026b is an extrasolar giant planet with an orbit of 2.87 days
around a metal-rich G0IV parent star. It has a radius of only
0.725R$_J$ $\pm$ 0.05R$_J$ and a mass of 0.36M$_{J}\pm$0.03M$_J$ (114
M$_{\oplus}$, \citealt{sato2005n2k}). Evolution models suggest that the
planet should have larger radii \citep{guillot1996giant}, but it is
decidedly small. \cite{fortney2006atmosphere} provide atmosphere
models for the planet and suggest that a hot stratosphere (temperature
inversion) may develop due to additional flux absorption by TiO and
VO. 
The (T$_{\rm gas}$, p$_{\rm  gas}$) profiles used in our study are shown in Fig.~\ref{TPall},
where $1\times$ (green) corresponds to solar abundance and $10\times$
(purple) corresponds to a 10 times solar atmospheric
profile. This model allows us to study the effect of an increase of element abundances by the same amount for each element compared to the solar element abundances. 

\paragraph*{Hot Super-Earth CoRoT-7b (HRSE):} 
We use (T$_{\rm gas}$, p$_{\rm gas}$) profiles from  \cite{Itoetal} for HRSE for a first assessment of cloud formation and  their potential details for CoRoT-7b should cloud formation be possible. The atmosphere simulation by
\cite{Itoetal} is a 1D plane-parallel thermal structure which is in
radiative, hydrostatic and chemical equilibrium. 
These models use the ground pressure, $P_g$ and molar fraction $x_A$
which are the functions of the ground temperature $T_{g}$ which is
determined assuming the magma ocean is a blackbody. No clouds are
considered in these models. Figure~\ref{TPall} shows several HRSE (T$_{\rm gas}$, p$_{\rm gas}$)-profiles provided by \cite{Itoetal} for hot super-Earths with 
T$_{\rm eq}$ from 2000 to 3000 K. In the cases of
$T_{\rm eq} \le$ 2000 K, the atmosphere is isothermal because
the atmosphere is so optically thin that the ground is directly heated
by the stellar irradiation \citep{Itoetal}.  \cite{Itoetal} used T$_{\rm equ}$=2300K to represent CoRoT-7b.

\paragraph*{Possible atmosphere on 55 Cnc e:} 55 Cnc e is an interesting candidate to study the potential atmospheric compositions due to its mass and radius estimated at 8.09 $\pm$ 0.26 M$_{\oplus}$ \citep{nelson201455} and 2.17 $\pm$ 0.10 R$_{\oplus}$ \citep{gillon2012improved} which makes it fall into the category of a Super-Earth. Given the extremely high equilibrium temperature (T$_{\rm eq}$ $\sim$ 2400 K) and its proximity to the parent star, it is highly likely that the planetary lithosphere is weak and most of the day-side surface of the planet will be in a semi-molten or molten state which would lead to magma oceans and possibly volcanic activity on the irradiated day-side \citep{demory2016variability}. 

Due to the high brightness of the host star, 55 Cnc e can be observed
with great detail in the visible as well as Spitzer 4.5 $\mu$m IRAC
Photometric band
\citep{demory2012detection,winn2011super,gillon2012improved}. Based on
the precise measurement of mass and radius alone,
\cite{demory2011detection} suggested a silicate-rich interior with a
dense H$_2$O envelope of 20$\%$ by mass, \cite{gillon2012improved}
suggested a purely silicate planet with no envelope and
\cite{madhusudhan2012c} suggested a carbon-rich planet with no
envelope, \cite{demory2016variability} suggested an atmospheric model
in which multiple volcanic plumes explain the large observed
temperature variations on the day-side. The (T$_{\rm gas}$, p$_{\rm
  gas}$) profile that we use in our study is derived from the figure
in \cite{demory2016variability}. This profile was retrieved based
  on the observed IRAC 4.5-$\mu$m brightness temperature (T$_B$)
  between T$_{min}$ = 1273$_{+271 K}^{-348 K}$ and T$_{max}$ =
  2816$_{+358 K}^{-368 K}$.  Such retrieved planetary atmosphere are
  therefore isothermal at the upper and the lower boundary, in
  contrast to complete forward models like {\sc Drift-Phoenix} or those
  from \cite{Itoetal} and \cite{fortney2006atmosphere}.

\subsection{Atmospheric mixing} 

The vertical mixing is parametrized using the eddy diffusion
coefficient K$_{zz}$ which represents the strength with which material
can be transported into higher atmospheric layers
\citep{agundez2014pseudo}. There has been extensive research on the
choice of K$_{zz}$ for the cases of hot-Jupiters
\citep{moses2011disequilibrium,miguel2013exploring,parmentier20133d,agundez2014pseudo}. \cite{parmentier20133d}
use vertical diffusive coefficients determined by following passive
tracers in their general circulation model (GCM). They adopt a
K$_{zz}$ [cm$^{2}$ s$^{-1}$] = 5 $\times$ 10$^{8}$ p $^{-0.5}$ bar for
HD 209458 b which is a gas-giant. \cite{agundez2014pseudo} used
K$_{zz}$ [cm$^2$ s$^{-1}$] = 10$^{7}$ p$^{-0.65}$ (bar) for the case
of HD 189733 b and \cite{miguel2013exploring} considered values between
K$_{zz}$ [cm$^{2}$ s$^{-1}$] =10$^{8}\ldots10^{9}$ for their
atmospheric models of mini-Neptunes.  We follow the approach detailed
in \cite{lee2015a} for the vertical mixing.  The diffusion mixing time
scale is derived from $ \mathrm{\tau_{mix} = const\cdot\,H_{\rm
    p}^2/K_{\rm zz}}$ for a prescribed diffusion constant K$_{zz}$.  A
constant K$_{zz}=10^{11}$ cm$^{2}$ s$^{-1}$ was chosen for 55 Cnc
e. Tests have shown that for the 55 Cnc e profile (black line joining
blue dots with error bars in Fig.~\ref{TPall}), cloud formation does
not start for K$_{zz}< 10^{9}$ cm$^2$ s$^{-1}$.  The K$_{zz}$ value
for HD 149026 b is selected to be 10$^{13}$ cm$^2$ s$^{-1}$.  For
CoRoT-7b, K$_{zz}$ was varied between values ranging from 10$^7$
cm$^2$s$^{-1}$ to 10$^{15}$ cm$^2$s$^{-1}$.



\begin{figure*}
\includegraphics[scale = 0.4]{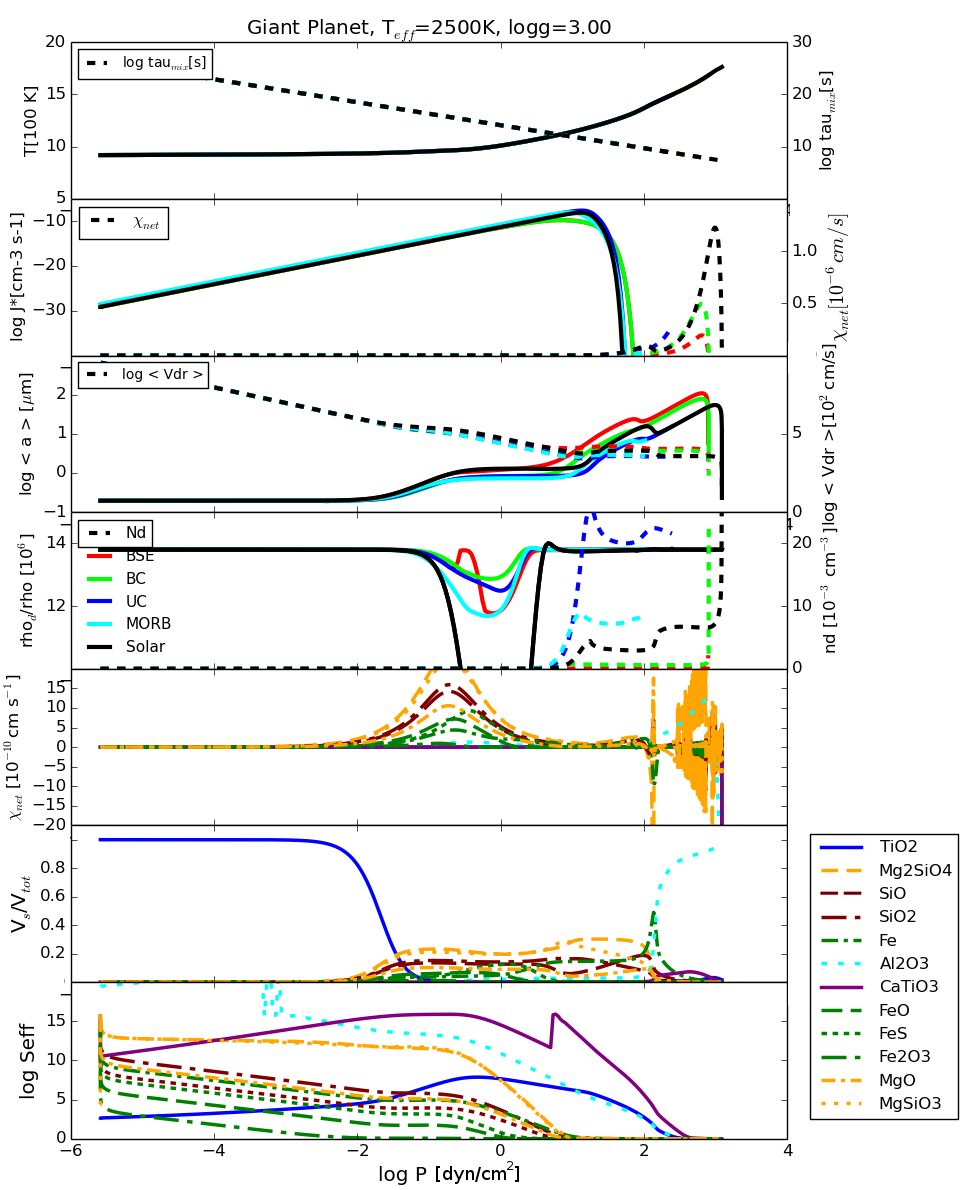}
\caption{Cloud structure and cloud details for a giant gas
  planet (T$_{\rm eff}$=2500K, logg=3.0) {\sc Drift-Phoenix}
  atmosphere model. \textbf{1st panel:} local gas temperature, T$_{\rm
    gas}$ (solid), mixing time scale $\tau_{\rm mix}$ (dashed);
  \textbf{2nd panel:} Nucleation rate, $\log J_*$ (solid), dust growth
  velocity, $\chi_{\rm net}$ ; \textbf{3rd panel:} mean cloud particle
  size, $\log \langle a\rangle$, drift velocity, $\log\langle v_{\rm
    dr}\rangle$; \textbf{4th panel:} dust-to-gas mass ratio,
  $\rho_{\rm d}/\rho_{\rm gas}$, particle number density, n$_{\rm
    d}$. The cloud properties based on five sets of initial element
  abundances (solar - black, BSE - red, Bulk Crust - green, Upper
  Crust - blue, MORB - sky blue) are shown in different colors in
  panels 2, 3, and 4. Panels 5 and 6 are for solar element abundances
  only, and the results are detailed for the 12 different growth
  species $s$: \textbf{5th panel:} grain growth velocity, $\chi_{\rm
    s}$; \textbf{6th panel:} material volume
  fraction, V$_{\rm s}$/V$_{\rm tot}$; \textbf{7th
    panel:} effective supersaturation ratio, $\log S_{\rm eff}$. }
\label{browndwarfcloud}
\end{figure*}

\begin{figure*}
\centering
\includegraphics[scale = 0.34]{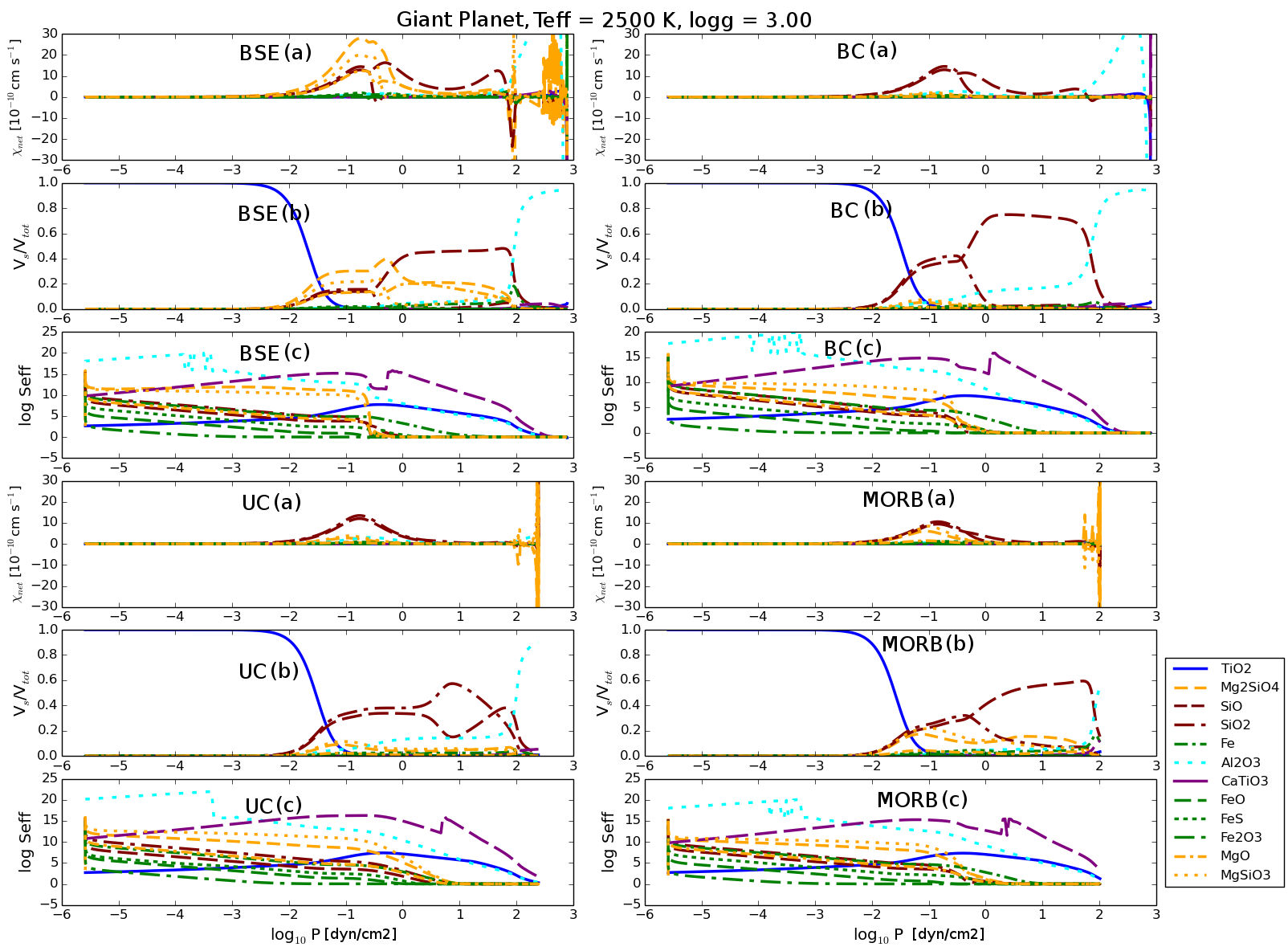}
\caption{The cloud particle properties results for a giant
  gas planet (T$_{\rm eff}$=2500K, logg=3.0) {\sc Drift-Phoenix}
  atmosphere model for four different sets of initial element
  abundances in detail for the 12 different growth species $s$:
  \textbf{(a)} grain growth velocity, $\chi_{\rm s}$ [cm s$^{-1}$];
  \textbf{(b)} material volume fraction, V$_{\rm s}$/V$_{\rm tot}$
         [\%]; \textbf{(c)} effective supersaturation ratio, $\log
         S_{\rm eff}$. The four compositions BSE, BC, UC, MORB have
         been labeled on each plot.}
\label{dustpropbd}
\end{figure*}

\begin{figure*}
\includegraphics[scale = 0.4]{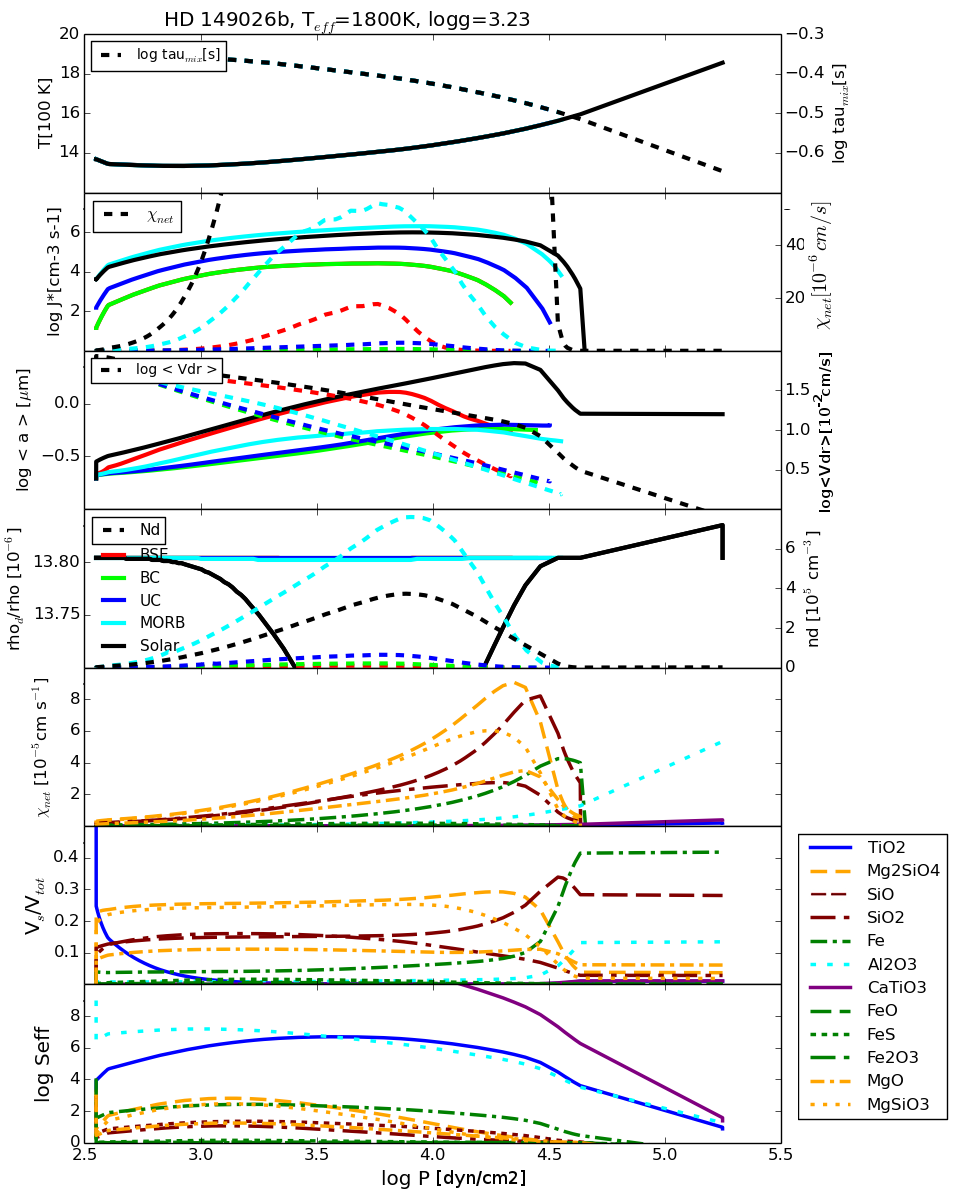}
\caption{Cloud structure and cloud details for HD149026b (T$_{\rm
    eff}$=1757K, logg=3.23) for five sets of initial element abundances
  (panel 2, 3, 4).  The figure has the same structure like
  Fig.~\ref{browndwarfcloud}.}
\label{cloudprophd149}
\end{figure*}

\begin{figure*}
\centering
\includegraphics[scale = 0.34]{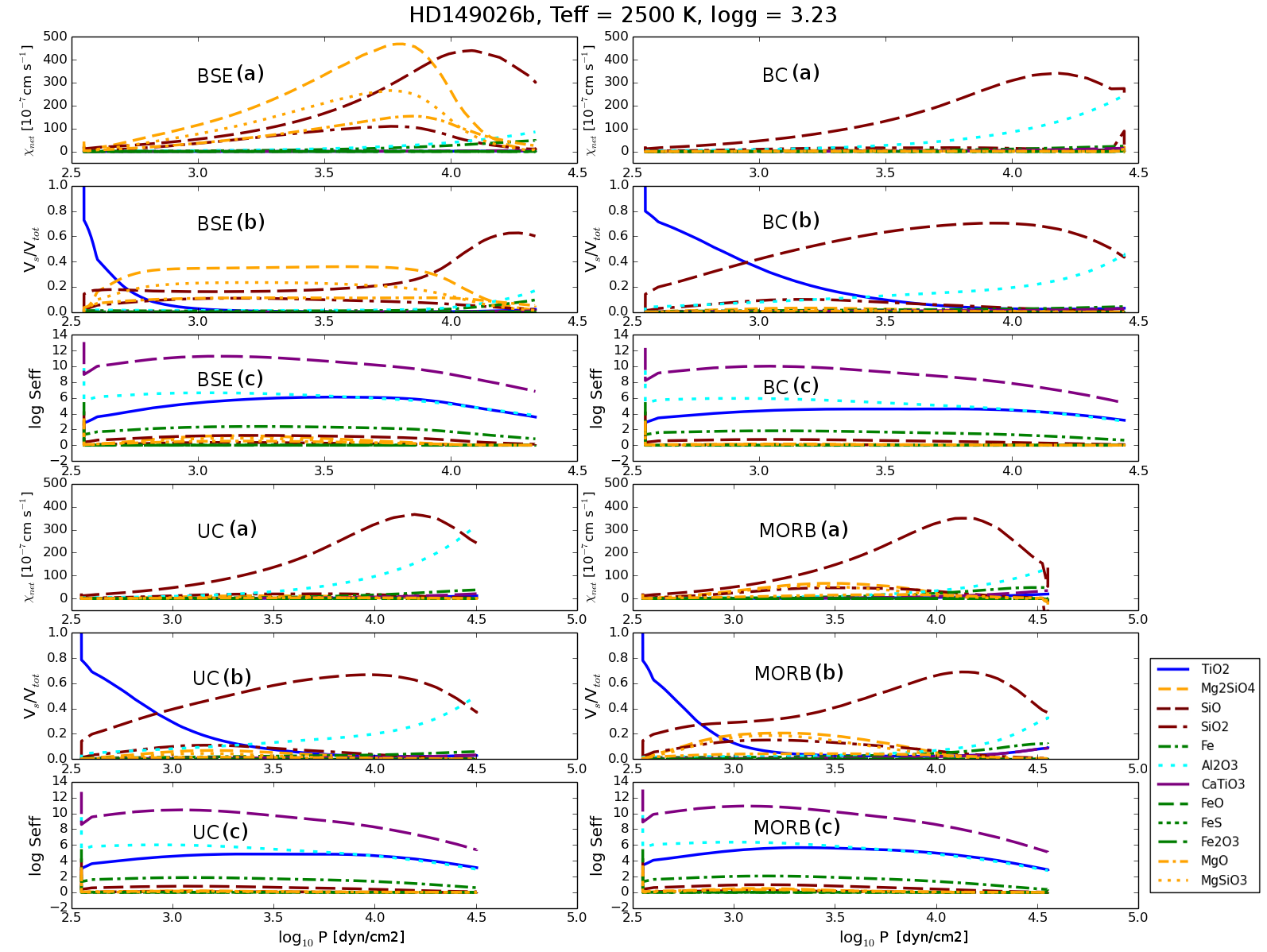}
\caption{Same like Fig.~\ref{dustpropbd} but for HD149026b (T$_{\rm
    eff}$=1757K,logg=3.23).}
\label{dustprophd149}
\end{figure*}

\begin{figure}
\hspace*{-0.8cm}
\includegraphics[scale = 0.28]{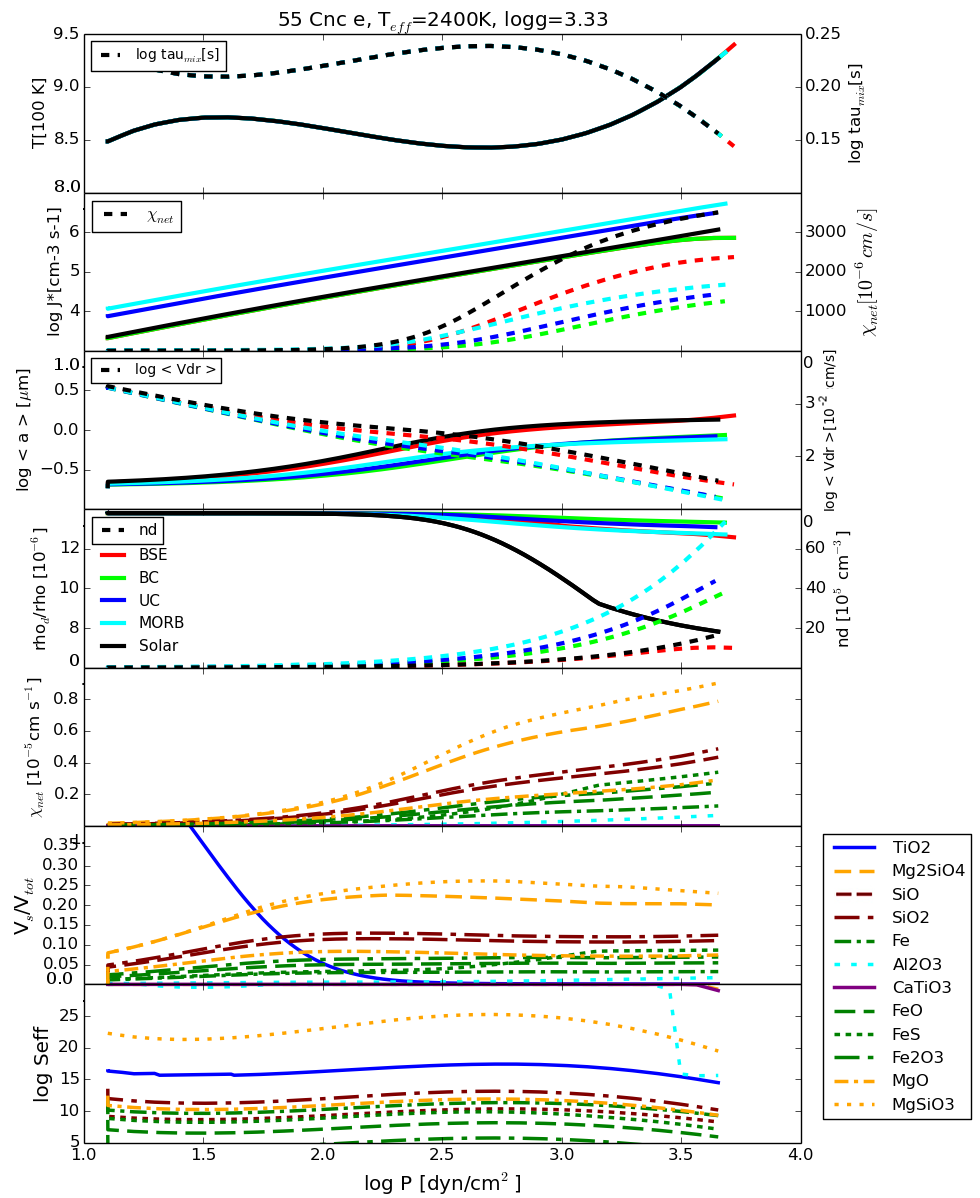}
\caption{Cloud structure and cloud details for for 55 Cnc e (T$_{\rm
    eff}$=2400K, logg=3.33) for five sets of initial element
  abundances (panel 2, 3, 4). The figure has the same structure like
  Fig.~\ref{browndwarfcloud}.}
\label{cloudprop55cnce}
\end{figure}

\begin{figure*}
\centering
\includegraphics[scale = 0.34]{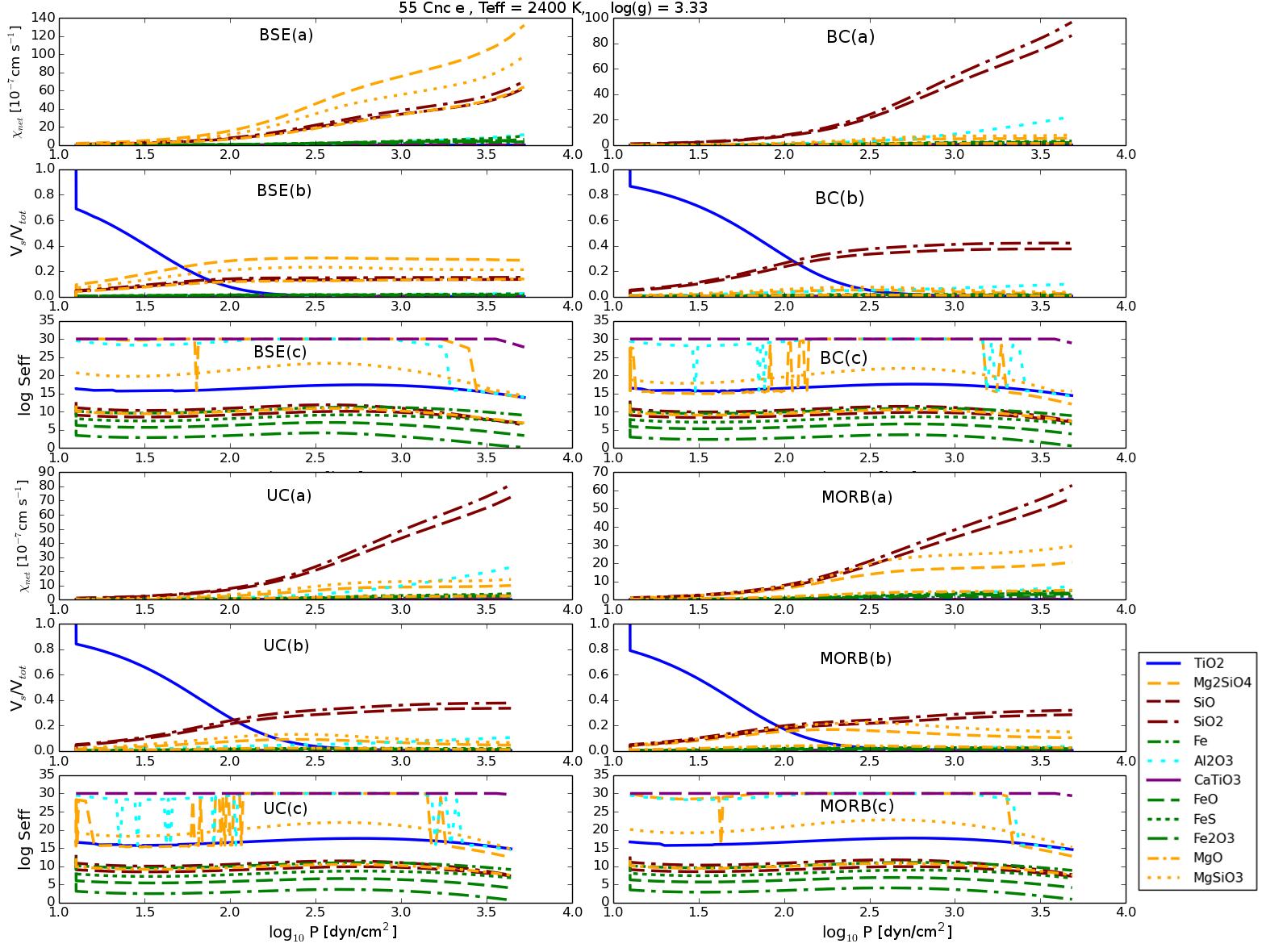}
\caption{Same like Fig.~\ref{dustpropbd} but for 55 Cnc e (T$_{\rm eff}$=2400K, logg=3.33).}
\label{dustprop55cnce}
\end{figure*}

\section{Results: Equilibrium gas-phase composition} \label{sec:gasphase}

Clouds form from an atmospheric gas if the appropriate thermodynamic
conditions are met. Their composition depends on the chemical
composition of the atmospheric gas. \cite{witte2009dust} demonstrated
how a (homogeneously) decreasing element abundance changes the cloud
structure in brown dwarf atmospheres by reducing the set of solar
element abundances by the same factor from
[M/H]=$0.0\,\ldots\,-6.0$. Recent chemical modelling efforts in planet
forming disks show that planets will most likely form from a set of
initial element abundances that is non-solar nor a simple scaling of
the solar set of element abundances (\citealt{hell2014,eis2016}). In
this section we show the results and analyze the chemical gas-phase
composition for all the cases explored in this paper. To keep the
number of panels sensible, we chose 4 sets of element abundances only
(solar, 10x solar, BSE and BSE with a low H abundance).  We note
  that all figure will also contain results for extreme cases for completeness and for reference. These
  extreme cases are the hydrogen-rich atmospheres for 55 Cnc e and
  CoRoT-7b which may only be realistic during a very short time of
  their infancy or when kicked back into a hydrogen-rich part of the
  planet-forming disk. The two BSE-cases for the giant gas planet and
  HD\,149026b can represent stages of a heavy influx of meteorits or
  even planetary collisions.

Figure \ref{Top10Gas} shows the 10 most abundant species that result
from our chemical equilibrium gas-phase calculations. In Figures
\ref{Si_All} to \ref{Ti_All} we show the results for the chemistry of
one element at a time (Fig. \ref{Si_All}: Si, Fig. \ref{Mg_All}: Mg
and Fe, Fig. \ref{Ti_All}: Al and Ti). We generally observe that a
lower hydrogen content (here: BSE (low H)) leads to higher abundances
of oxides such as SiO$_2$, TiO$_2$, MgO, FeO, AlO$_2$H.

\paragraph*{Giant gas planet:}

The atmosphere is dominated by volatiles for the solar, metal enriched
solar (10x solar) and BSE cases. In the case of solar abundance, the
atmosphere is primarily composed of H$_2$, H, CO, as shown in
Figure~\ref{Top10Gas}.  Increasing the element abundances
homogeneously results in an increase of the  molecular number
  densities (compare 1st and 2nd column in Figs.~\ref{Si_All} --
~\ref{Ti_All}). This effect is stronger for Mg, Si and Fe than for
elements with an intrinsically lower abundance like Ti and Al.

The BSE atmosphere has an increase in H$_2$O, SiO, SiS, K, Al, TiO among others (see Figure~\ref{Top10Gas}). The differences in the solar and
BSE abundances with a fixed C/O ratio of 0.5 is quite pronounced in
the higher pressure regions where the amount of TiO and TiO$_2$
increases slightly with a drastic decrease in TiC molecules in the BSE case. An
increase in such UV absorbers may therefore induce a local temperature
inversions \citep{fort2006}.  The strongest changes with metallicity
occurs for the Mg-binding species. 

\paragraph*{55Cnc e:} 
The atomic species dominate the gas-phase for all elements in the case
of BSE composition. \cite{tsiaras2015detection} in their analysis of
an atmosphere around 55 Cnc e, show that the abundances of HCN and
C$_2$H$_2$ increases while the abundance of H$_2$O decreases
drastically which is similar to the results that we obtain for a test
BSE composition with C/O=1.0  (not shown) as compared to a
oxygen-rich solar abundance.

\paragraph*{Metal-rich mini-giant  planets, e.g. HD149026b:} The third rows of Figs.~\ref{Si_All}, \ref{Mg_All} and \ref{Ti_All} show the gas-phase composition for HD149026b. The temperature inversion is seen at around 2700 K where all the gases go through a sudden increase in concentration. The second panel of the third row for the figures show the plots for an enriched atmosphere with 10x the solar metallicity.  The abundance of oxides of Ti, Al, Fe, Mg and Si is higher for
HD149026b as compared to other planets which would add to the cloud
opacity by imposing strong cloud particle growth.

\paragraph*{Hot Super-Earth CoRot 7b:}  
The fourth row of Figs. \ref{Si_All} -- \ref{Ti_All} 
show the number densities of the dominating species for CoRoT-7b case.
The local gas temperatures for this planet are considerably higher
compared to the other three planetary model atmospheres due to the
stronger irradiation and the hight temperatures profiles adopted from \cite{itoetal}. The atmospheric model of CoRoT-7b is
representative of a rocky planet. Due to the closeness to its host
star and high irradiation, the volatiles might have escaped. Therefore
we would expect the BSE case with a low hydrogen content to be
the most representative of CoRoT-7b initial element abundances. A
comparison between BSE case with the solar element abundances shows
the sensitivity of the gas composition to uncertainties in initial
element abundances. We note that a changing hydrogen-content will
  affect the mean molecular weight of the whole atmosphere.

The atmosphere of CoRoT--7b is primarily composed of gases in their
atomic and ionic states, hydrogen is present as H (see
Fig.~\ref{Top10Gas}).  Si species such as SiO and SiS are still quite
dominant. We find most of the other elements such as Ti, Mg, Fe and Al
to be dominant in their atomic and ionic states in both solar and BSE
atmospheres. SiO dominates the Si-binding gas species at high gas
pressures, followed by Si and Si$^+$. This hierarchy is somewhat
metallicity dependent. Mg, Fe, Al and Ti appear always in their atomic
states followed by their singly ionised state (Mg/Mg$^+$, Fe/Fe$^+$,
Al/Al$^+$, Ti/Ti$^+$). Al$^+$ and Ti$^+$ appear at lower pressures
than Mg$^+$ or Fe$^+$ for the CoRoT-7b profile utilised
here. Therefore, the atmosphere of CoRoT-7b is strongly ionised on its
day-side. This suggests that either no cloud formation can take place
or other cloud formation paths than those considered here need to be
considered. For example, noctilucent clouds on Earth are suggested to
form through ion-cluster nucleation.

A decreasing hydrogen abundance does not affect these results significantly with respect to the most abundant species of each element.

\subsection{Summary}

We have analyzed the gas-phase composition for different set of initial
abundances as starting point for cloud formation. The
primary abundant species is H$_2$ in all of our atmosphere's except
that of CoRoT-7b which has atomic H as the dominating species due to
its higher gas temperatures. The BSE atmosphere model used for CoRoT-7b
suggests the presence of gases such as SiO, Mg, K, Na, O which are
also found in the atmospheres of 55 Cnc e and HD149026b. SiO is the
dominant gas resulting from Si in all of the planet cases except
CoRoT-7b where atomic Si is the dominant gas-phase species due to
higher local gas temperatures (except at high gas pressures). It is
followed by SiO$_2$ in the upper parts of the atmospheres for the gas
giant, 55 Cnc e and HD149026b.

Atomic Fe is the dominant species among the Fe-binding gas-phase
species, for all of the planets and element compositions considered
here. Similarly, atomic Mg is the dominant species among all of the Mg
bearing gas-phase molecules, except for HD149026b with BSE (low H)
where we see MgO as another dominant species.  However, giant
  planets with low hydrogen content are an extreme and most likely
  unrealistic case.  Our results suggest that the chemical gas-phase
composition does not change drastically if the element abundances are
increased by the same factor for all elements starting from solar
composition (up-scaled solar composition), but it leads to higher
gas-phase concentrations of enriched elements (as can be seen in the
second column of Figs.~\ref{Si_All}, \ref{Mg_All}, \ref{Ti_All}).  We
note that our gas-phase results for CoRot-7b so far suggest that the
atmosphere is highly ionised and plasma process may therefore effect
the atmosphere (\citealt{2016SGeo}).

\section{Clouds in atmospheres of non-solar element abundances} \label{sec:clouds}

Table~\ref{planetschar} summarizes the global parameters of the planets that we use to evaluate cloud
structures for planets that have been suggested to possess
potentially non-solar sets of initial element abundances. 

Figures~\ref{browndwarfcloud},~~\ref{cloudprophd149}
and~\ref{cloudprop55cnce} show the input (T$_{\rm gas}$, p$_{\rm
  gas}$, $\tau_{\rm mix}$) in the top panel. Panel 2 -- 4 demonstrate
governing cloud variable (nucleation rate $J_*$ [cm$^{-3}$ s$^{-1}$],
  net growth velocity $\chi_{\rm net}$ [cm/s], equilibrium drift
  velocity v$_{\rm dr}$ [cm/s]) and resulting cloud properties (cloud
  number density n$_{\rm d}$ [cm$^{-3}$ ], mean cloud particle size
  $\langle a\rangle$ [$\mu$m], dust-to-gas ration $\rho_{\rm
    d}/\rho_{\rm gas}$) for 5 sets of element abundances (BSE, BC, UC,
  MORB, solar). Panels 5, 6, and 7 present the results for the set of
  solar element abundances only but now detailed for the 12 individual
  growth materials involved (see Sect.~\ref{ss:cloudinp}): net growth
  velocity $\chi_{\rm net}^{\rm s}$ [cm/s] for material $s$, the
  volume fraction $V_{\rm s}/V_{\rm tot}$ for material $s$ and the
  effective supersaturation ration S$_{\rm eff}$ for each material
  $s$.  Figures~\ref{dustpropbd},~\ref{dustprophd149}
  and~\ref{dustprop55cnce} contain 4 groups of three panels that are
  similar to Panels 5, 6, and 7 in
  Figs.~\ref{browndwarfcloud},~\ref{cloudprophd149}
  and~\ref{cloudprop55cnce} but for the remaining sets of element
  abundances (BSE, UC, BC, MORB).

\subsection{Clouds that form in the atmosphere of a giant gas planet}\label{ss:ggp}

 Figure \ref{browndwarfcloud} shows how the cloud properties and, hence, the cloud structure changes in the case of different sets of initial element abundances for a sample atmosphere representative of a non-irradiated gas-giant (T$_{\rm  eff}=$2500 K, log(g)= 3.0) being representative for a long-period or free-floating planet.

 The nucleation rates (J$_{*}$, panel 2, left) appear similar for the
 sets of element abundances, however bulk-crust (BC) abundance causes a less efficient
 nucleation at the nucleation peak. BC further allows the nucleation
 to spread somewhat more towards higher temperatures and higher gas
 pressure than any of the other sets of element abundances. The net
 growth velocity ($\chi_{\rm net}$, panel 2, right) reaches largest
 values for solar abundance. There is significant variation in the
 final particle sizes ($\langle a\rangle$, panel 3, left) before the
 cloud particles evaporate. The particles for the
 BSE set of initial element abundances reach the biggest sizes
 followed by the BC, solar and then MORB and upper-crust (UC) abundance sets.  This
 is a clear hierarchy of available gas-phase growth species rather
 than an effect of the local gas temperature because no  detailed  feedback onto
 the gas temperature is taken into account in our present work  based on the new cloud structures derived here.  The underlaying model atmospheres from {\sc Drift-Phoenix} and  from \cite{fortney2006atmosphere} do include the effect of clouds on the temperature structure and element abundances.   This is also seen
 from the drift velocities ($v_{dr}$, panel 3, right) with BSE
 particles being the biggest and having the largest drift velocity.
 The individual material volume fractions ($V_{\rm s}/V_{\rm tot}$,
 panel 6) reflect the individual net growth velocities ($\chi_{\rm
   net}^{\rm s}$, panel 5) of each of the materials, $s$, and
 demonstrates where growth is most efficient for which material. The
 composition for the solar set of element abundances is dominated by
 TiO$_2$[s] seed formation in the upper atmosphere.  Other dust
 species start to condense on the seed decreasing its volume fraction
 steadily. A large fraction of the cloud layer is dominated by
 Mg/Si/O-containing materials and below that (i.e. at higher
 temperatures) Al- and Fe-containing species occupy a larger volume
 fraction by forming predominantly Al$_2$O$_3$[s] and Fe[s] until the
 particles become thermally unstable and evaporate completely. More
 details can be found in earlier publications
 (e.g. \citealt{helling2008,helling2014atmospheres,lee2015modelling,hell2016}).

 Figure \ref{dustpropbd} provides details about the variation of material composition
 of the cloud particles for non-solar sets of element abundances.  We see an atmosphere with BSE  composition to be consisting of Mg-binding materials ($\sim$ 30$\%$),
Si-binding materials ($\sim$ 40 $\%$)
and finally Al$_2$O$_3$[s] dominates as the primary dust species until
a pressure of 10$^{3}$ dyn/cm$^2$ before the particles evaporate. The
atmospheres with a BC, UC and MORB type of compositions have
significant differences as compared to BSE, with the cloud particles
being primarily composed of SiO[s] and SiO$_2$[s] which
in a combined manner constitute of more than 80$\%$ of the cloud volume
fractions in each case. Hence, objects with  a MORB-type (or UC/BC) set of initial element abundances can be expected to have mineral clouds predominantly made up of Si-O cloud particles with smaller impurities from high-temperature condensates.

\subsection{Clouds that form  in the atmosphere of the metal-rich giant gas planet HD149\,026\,b}

HD149\,026\,b can be classified as a mini-giant planet with a mass of
  114 M$_\oplus$. \cite{fortney2006atmosphere} investigate the
  atmospheric and cloud properties for the planet with varying
  metallicities such as [M/H] of 1x , 3x , 10x with TiO and VO
  enrichment for 3x and 10x. We perform our cloud modelling on a
  similar (T$_{\rm gas}$, p$_{\rm gas}$) profile for 1x Solar
  abundance as shown in the Fig.~\ref{TPall} while varying the
  silicate abundances according to our four different Earth silicate
  compositions of BSE, BC, MORB and UC. Our models suggest a high
  value of vertical mixing with K$_{zz}=10^{13}$ cm$^2$s$^{-1}$ or
  higher is needed for the atmosphere to be able to form dust clouds.

Figures \ref{cloudprophd149} and \ref{dustprophd149} show the dust
cloud properties for HD149\,026b as result of our kinetic cloud
formation model. Also here, the cloud formation is triggered by the
occurrence of TiO$_2$ nucleation. Similar to previous findings or the
giant gas planet atmosphere model, the nucleation rates ($J^*$, panel
2, left) differs by 1-2 orders of magnitude for the different sets of
non-solar element abundances. The MORB composition produces the
highest nucleation due to its higher Ti abundance (see Table
\ref{tab:EAb}). The particle growth velocities are the highest for a
solar composition and the particles reach the largest sizes ($\sim$ 3
$\mu$m) followed by a BSE composition. The drift velocities ($\langle
v_{dr}\rangle$) are found to be in the same sequence of solar followed
by BSE, MORB, BC and UC respectively. The particles formed from a
solar composition gas are thermally stable in a larger range of
pressure before they evaporate completely. The dust number densities
(n$_{\rm d}$) are the highest for a MORB composition again as
consequence of the highest $J_*$.

  A solar and BSE composition in HD149026b follows a similar trend of
  dust composition wherein, just after the nucleation, Mg-binding silicates
  dominate the dust volume fraction with $\sim$40$\%$ of the dust
  having Mg-binding dust species and is followed by Si- ($\sim$40$\%$) and
  Fe-($\sim$30$\%$) binding species which dominate the cloud base. For
  BC, UC and MORB compositions, our models suggest that the cloud particles
  majorly are made of Si-binding species ($\sim$ 60 $\%$) and the cloud
  base having an increased amount of Al$_2$O$_3$[s] which increases up to
  $\sim$40$\%$ at the cloud base. Table \ref{dustvolfractions} shows
  the average volume fractions for each of the dust forming species on
  the three of the possible atmospheres.
 \cite{fortney2006atmosphere} consider
  atmospheric inversions due to the presence of TiO in their models
  but our dust cloud models suggest an upper atmosphere in which TiO
  is heavily depleted due to the seed formations.  This is not a new finding and has been demonstrated for many cases (e.g. \citealt{helling2008,witte2009dust,lee2015modelling}). Such an element depletion  would result in
  a loss of temperature inversion or suppression of the inversion zone
  to lower parts of the atmosphere where the TiO density is higher.

The cloud structure for HD149026b is comparable to the results for the
hot giant gas planet (Sect.~\ref{ss:ggp}), but the cloud reaches
considerably deeper into the atmosphere.

\subsection{Clouds in the atmosphere of 55 Cnc e}

Figures \ref{cloudprop55cnce} and \ref{dustprop55cnce} show the
cloud properties for 55 Cnc e. Cloud formation only occurs in a thin
atmospheric region near the top of the adopted (T$_{\rm gas}$, p$_{\rm
  gas}$) profile where the local gas temperature and densities allow
seed formation and efficient subsequent particle growth (compare
pressure interval in Fig.~\ref{cloudprop55cnce} and
Fig.~\ref{TPall}). The extension of the cloud layer in the atmosphere
of 55 Cnc e is therefore considerably smaller compared to the other
cases studied here.

TiO$_2$-nucleation occurs throughout the whole atmosphere domain
available for 55 Cnc e. Seed formation (J$_*$, panel 2, left) differs
by about one order of magnitude for the different sets of element
compositions, which effects the number of cloud particles formed
accordingly. The MORB composition attains the highest nucleation rates
which can be attributed to its higher abundance of Ti molecules
leading to efficient nucleation. The sets of solar and BC element
abundances results in the least efficient nucleation rate, hence, less
cloud particles, n$_{\rm d}$, are formed.  The growth velocities
(panel 2) are the highest for a solar composition followed by BSE, BC,
UC and MORB. The particles encounter denser regions as they settle
gravitationally and thereby also increase their size in the process
due to availability of more reaction material. In the case of 55 Cnc
e, we observe a smooth change with height of the cloud material
composition which is due to the fact that the cloud layer is thin as
compared to the cloud layer in the gas giant. This, however, maybe a
bias due to the domain of the prescribed atmosphere model available.
The Solar and BSE compositions have Mg-binding silicates
($\sim$25$\%$) for the major part of the dust cloud and it further
increases to $\sim$30$\%$ of the dust before
evaporation. Si-containing growth species SiO and SiO$_2$ constitute
$\sim$15$\%$ of the cloud particles. Fe, Al and Ca species are present
in minority ($<$5$\%$) in BSE atmosphere. SiO and SiO$_2$ form the
majority of the dust particles for BC, UC and MORB atmospheres which
is due to lower Mg content available to condense on the dust
particles.

Our results for 55 Cnc e demonstrate how sensible cloud formation
reacts  on the initial set of element abundances: solar abundances
disfavour an efficient seed formation (based on TiO$_2$) but allow for
the most efficient surface growth processes among the sets of element
abundances studied here. In contrast, also BC disfavours efficient
nucleation but it also slows down surface growth being the least
efficient in the sets studied. Consequently, the solar case produces
clouds with few but big particles which rapidly rain out, and the
cloud particles forming based on other sets remain suspended in the
atmosphere for longer due to their smaller sizes.

\subsection{The possibility of clouds in the atmosphere of CoRoT-7b}

{CoRoT-7b is an interesting candidate to analyze the possibility of
  cloud formation due to its extremely high local gas temperatures
  ($>$2500 K) and low local gas pressures (-2 $< \log(p_{\rm gas}) <$
  2 [dyn/cm$^2$]) making it a challenging candidate for cloud
  formation based on the presently available (T$_{\rm gas}$, p$_{\rm
    gas}$)-structures. Figure~\ref{TPall} demonstrates the large
  differences between those atmospheres where we have shown that
  clouds form and the atmosphere structures proposed for HRSE objects
  like CoRoT-7b.  We investigated if cloud formation is possible on
  CoRoT-7b's day-side by using the sample atmosphere profiles for a
  rocky planet with four different T$_{\rm eq}$ (Fig.~\ref{TPall}).
  Our kinetic cloud model does not suggest any cloud formation on the
  day-side of CoRoT-7b for the atmosphere structures provided. Our
  finding, however, does not preclude cloud formation on the night
  side.

  \cite{schaefer2009} predict clouds of Na and K in the planet's
  atmosphere based on their phase-equilibrium approach. However,
  \cite{gao2016} show that it can not be taken for granted that such
  species actually nucleate, and argument that has been put forward in
  \cite{hell2013}. Phase-equilibrium approaches focus on possible
  end-results of condensation processes but do not offer insight into
  the process of formation (\citealt{hell2008c}). The (T$_{\rm gas}$,
  p$_{\rm gas}$)-structures utilized here for CoRoT-7b appear hot
  enough that a strong ionisation of the atmospheric gas should be
  expected according to our gas-phase composition study in
  Sect.~\ref{sec:gasphase}. Given the high irradiation  onto an existing atmosphere, strong winds
  will develop and transport ionised gas into cooler atmospheric
  regions where clouds will form similar to 55 Cnc e. The
  night-side/day-side transition region will most likely be
  electrostatically active and electric currents are likely to affect
  the weather on CoRot 7b as suggested in e.g. the review by
  \cite{hell2016}. 


\subsection{Summary}

Table~\ref{table:sum} provides an overview of global changes of the
cloud structure depending on the sets of initial element abundances. We
chose to visualise this in terms of volume fractions (V$_{\rm
  s}$/V$_{\rm tot}$ [$\%$]) at three typical heights inside the cloud,
the maximum nucleation rate, and the average mean grain size
throughout the 1D cloud layer.  These values can serve as guidance for retrieval methods.

The highlighted values for V$_{\rm s}$/V$_{\rm tot}$ visualise that
several materials dominate the cloud particle composition but the type
of material chances with height in all cases
studied. Table~\ref{table:sum} demonstrates further that a number of
materials do contribute with less than 10\% to the overall cloud
particle volume (CaTiO$_3$[s], Fe$_2$O$_3$[s], FeS[s], FeO[s]).

The nucleation species TiO$_2$[s] dominates the cloud particle
composition at the cloud top while Mg-Si-O materials provide the
material bulk independent on the set of initial element abundances. 

We use the maximum nucleation rate and the average mean grain size as
measured to compare our results for the individual objects (hot gas
giant, HD\,149026b, 55 Cnc e). This clearly shows that the effect of
the different initial element abundances are negligible compared to
differences in local thermodynamic conditions. The anti-correlation
between nucleation rate (hence, number of cloud particles formed) and
mean grain size outlined earlier prevails also globally.

CoRot7b has no clouds forming on its day side and a strong
day-side/night-side contrast should therefore be expected which is
determined by the planet's cloud formation.

\hfill \break
\hfill \break

\begin{table*}
\centering

\caption{A comparison of characteristic cloud properties for a 
  gas giant, 55 Cnc e and HD\,149026b: the volume fractions (V$_{\rm
    s}$/V$_{\rm tot}$ [$\%$]) for individual materials $s$, the
  maximum nucleation rates ($\log_{10}$J$_{\rm *, max}$ [cm$^{-3}$s$^{-1}$] ) and cloud-averaged mean cloud particle radius ($\langle a\rangle_{\rm avg}$ [$\mu$m]) for three types of compositions. The cloud properties are compared for the cloud top, the middle (approximate half-length of the cloud) and for the
  cloud base (where the cloud particles have
  evaporated). All V$_{\rm  s}$/V$_{\rm tot}>10$\% are shown in boldface.} \label{dustvolfractions} \hfill \break
\resizebox{\textwidth}{!}{%
\begin{tabular}{c c | c c c | c c c | c c c}

    \hline \hline
    {\bf Volume fraction $V_{\rm s}/V_{\rm tot}$} [$\%$] & $\epsilon_{\rm i}^0$  & \multicolumn{3}{c}{\textbf{Gas Giant}}
    & \multicolumn{3}{c}{\textbf{55Cnc e}} & \multicolumn{3}{c}{\textbf{HD149026b}}\\
    \hline
     & & Cloud Top & Middle & Base & Cloud Top & Middle & Base & Cloud Top & Middle & Base\\
     p$_{\rm gas}$ [bar]& & 10$^{-5.5}$  & 10$^{1.3}$ & 10$^{3}$ & 10$^{6.04}$ & 10$^{6.4}$ & 10$^{6.7}$ & 10$^{2.5}$ & 10$^{3}$ & 10$^{5.2}$\\
     \hline
     & Solar & {\bf 100.0} & 0.04 & 1.66 & 0.4 & 0.1 & 5.05 & 99.6 & 2.2 & 1\\
    TiO$_2$[s]   & BSE & {\bf 100.0}  & 0.1  & 4.47 & 1.67 & 0.13 & 6.4 & {\bf 99.9} & 4.61 & 2.2\\
     & BulkCrust & {\bf 100.0} & 0.3 & 5.8 & {\bf 29} & 0.9 & 9.9 & {\bf 99.9} & {\bf 45.2} & 3.02\\
    \hline
     & Solar & 0 & 2.79 & {\bf 97.4} & 0.7 & 1.07 & 3.9 & 0.002 & 0.8 & 13.4\\
    Al$_2$O$_3$[s]   & BSE & 0 & 6 & {\bf 94.2} & 0.9 & 1.35 & 6.13 & $<$0.001 & 1.2 & 17\\
     & BulkCrust & 0 & {\bf 17.7} & {\bf 93.1} & 3.6 & 6.3 & 16 & $<$0.01 & 6.7 & {\bf 45}\\
    \hline
    & Solar & 0 & 0.15 & 0.47 & 0.01 & 0.01 & 0.35 & $<$0.001 & 0.05 & 1.1\\
    CaTiO$_3$[s]   & BSE & 0 & 0.16 & 1 & 0.03 & 0.05 & 0.35 & 0 & 0.07 & 1.0\\
     & BulkCrust & 0 & 0.7 & 0.7 & 0.13 & 0.16 & 0.96 & 0 & 0.4 & 2.6\\
    \hline
     & Solar & $<$0.001 & $<$0.001 & $<$0.001 & 5.15 & 6 & $<$0.001 & $<$0.001 & $<$0.001 & 0\\
    Fe$_2$O$_3$[s]   & BSE & 0 & 0 & 0 & 0.3 & 0.33 & 0 & 0 & 0 & 0\\
     & BulkCrust & 0 & 0 & 0 & 0.08 & 0.14 & 0 & 0 & 0 & 0\\
    \hline
     & Solar & 0 & 0.08 & $<$0.01 & 4.8 & 8.02 & 0.2 & 0.01 & 1.4 & 0.1\\
    FeS[s] & BSE & 0 & 0.05 & 4.2 & 1.7 & 2 & 0.1 & $<$0.001 & 0.15 & 0.04\\
     & BulkCrust & 0 & $<$0.01 & 0 & 1.36 & 1.86 & 0.02 & 0 & 0.04 & $<$0.01\\
    \hline
     & Solar & 0 & $<$0.01 & $<$0.01 & 5.6 & 5.6 & 0.2 & 0.01 & 0.5 & 0.05\\
    FeO[s]   & BSE & 0 & 0 & $<$0.001 & 1.11 & 1.25 & 0.02 & $<$0.001 & 0.01 & $<$0.01\\
     & BulkCrust & 0 & 0 & 0 & 0.85 & 1.16 & $<$0.01 & 0 & $<$0.01 & $<$0.01\\
    \hline
     & Solar & 0 & {\bf 14.6} & 0.34 & 3.3 & 3.3 & {\bf 18.1} & 0.01 & 4 & {\bf 41.7}\\
    Fe[s]   & BSE & 0 & 4.11 & 0.14 & 0.6 & 0.7 & 4 & 0 & 0.8 & 9.5\\
     & BulkCrust & 0 & 2.2 & 0.05 & 0.5 & 0.7 & 1.9 & 0 & 0.8 & 4.2\\
     \hline
     & Solar & 0 & 9.7 & 0.08 & {\bf 12.4} & {\bf 11.7} & {\bf 46.4} & 0.05 & {\bf 15} & {\bf 28}\\
    SiO[s] & BSE & 0 & {\bf 45.9} & 0.16 & {\bf 13.4} & {\bf 13.4} & {\bf 66.4} & $<$0.001 & {\bf 16.2} & {\bf 60.1}\\
     & BulkCrust & 0 & {\bf 72.3} & 0.16 & {\bf 25} & {\bf 36.6} & {\bf 69.8} & $<$0.01 & {\bf 35.6} & {\bf 44.5}\\
    \hline
     & Solar & 0 & {\bf 14.8} & $<$0.001 & {\bf 14} & {\bf 13.2} & 3.5 & 0.05 & {\bf 15.7} & 2.8\\
    SiO$_2$ & BSE & 0 & 1.01 & $<$0.001 & {\bf 15.1} & {\bf 15} & 2.32 & $<$0.01 & {\bf 10.0} & 2.0\\
     & BulkCrust & 0 & 3.14 & $<$0.01 & {\bf 28.1} & {\bf 41.1} & 0.8 & $<$0.01 & 7.4 & 0.5\\
    \hline
     & Solar & 0 & 5.1 & $<$0.001 & 9.1 & 8.35 & {\bf 10.5} & 0.03 & {\bf 11} & 6.1\\
    MgO[s] & BSE & 0 & {\bf 15.8} & $<$0.001 & {\bf 12.3} & {\bf 12.5} & 9 & $<$0.01 & {\bf 10.5} & 5.0\\
     & BulkCrust & 0 & 0.6 & $<$0.01 & 1.07 & 1.05 & 0.18 & 0 & 0.5 & 0.07\\
    \hline
     & Solar & 0 & {\bf 22.4} & $<$0.001 & {\bf 23.1} & {\bf 23.1} & 3.2 & 0.07 & {\bf 24.2} & 2 \\
    MgSiO$_3$[s] & BSE & 0 & 6.05 & 0 & {\bf 22.4} & {\bf 22.7} & 1.9 & $<$0.01 & {\bf 22.1} & 1.3\\
     & BulkCrust & 0 & 0.6 & 0 & 6.0 & 5.9 & $<$0.01 & $<$0.01 & 1.0 & $<$0.01\\
    \hline
    & Solar & 0 & {\bf 30.2} & $<$0.001 & {\bf 21.3} & {\bf 19.4} & 8.3 & 0.07 & {\bf 25.4} & 3.6\\
    Mg$_2$SiO$_4$[s] & BSE & 0 & {\bf 20.7} & 0 & {\bf 30.1} & {\bf 30.3} & 3.3 & $<$0.01 & {\bf 34.1} & 1.6\\
     & BulkCrust & 0 & 2.2 & 0 & 4.17 & 4.1 & 0.13 & $<$0.01 & 2.06 & 0.05\\
    \hline
   \hline
   \textbf{Maximum nucleation rate}     & Solar &  & -8.1 &  &  & 12.13 &  &  & 5.98 & \\
    log$_{10}$J$_{\rm *, max}$ [cm$^{-3}$s$^{-1}$] & BSE & & -9.75 & &  & 11.86 & & & 4.42 & \\
      & BulkCrust & & -8.91 & & & 12.35 & & & 4.74 & \\
     \hline
    \textbf{Cloud-averaged mean cloud particle radius}   & Solar &   & 2.83 &   &  & 0.50 &   &  & 0.58 & \\
    $\langle a\rangle_{\rm avg}$ [$\mu$m] & BSE &  & 3.58 &  &  & 0.47 &  &  & 0.43 & \\
      & BulkCrust &  & 2.98 &  &  & 0.32 &  &  & 0.29 & \\
    \hline
    \end{tabular}}
    \label{table:sum}
\end{table*}
\hfill \break


\section{Conclusions}\label{sec:summary}

Element abundances are important input properties for atmosphere (and
evolutionary) modelling which are, however, rarely known.  Determining
element abundances for extrasolar planets would be a valuable
exercise as this might allow to back-trace their place inside a
planet-forming disk. However, chemical processing inside the protoplanetary disk and
in the planet's atmosphere need to be known well enough for this
exercise. 

 {This work is the first attempt to model cloud
  formations in planetary atmospheres with Earth-silicate like
  element compositions.   Our equilibrium chemistry gas-phase composition
   reflects the adopted silicate element compositions from the rocky and the solar element abundances. Apart from the
  volatile rich atmosphere consisting primarily of H$_2$, H and CO
  there is an increase in content of gases such as Mg, Fe, SiS, Al,
  Ca, Si, Ti and K. 
  
  Our results suggest that clouds form in the atmospheres of 55 Cnc e, 
  HD149\,026b, and also in the hot gas giant as we demonstrated in previous works.  Cloud formation in atmospheres of evaporating magma planets (referring to planets with a molten surface)  like 55 Cnc e requires a vertical recirculation of condensible material of $>$ 10$^{11}$ cm$^{2}$s$^{-1}$. Vertical replenishment mechanisms have not been explored in detail in this
  work but a large day-night temperature gradient may favour a
  circulation mechanism and hence increase the likeliness of 
  cloud formation at or near the sub-stellar point. Such possibilities have not been ruled out by \cite{demory2016map} wherein the large observed temperature gradient between the day and night side may indeed favour cloud formation. They also find the hotspot to be shifted by 41 $\pm$ 12 degrees towards the east of the sub-solar point which further augments the possibility of a strong circulation regime. Our cloud model indicates the possibility of a highly opaque cloud layer on 55 Cancri-e with  particles primarily consisting of Mg silicates
  ($\sim$25$\%$), followed by Si-oxides and a minor percentage of Fe-, Al- and
  Ca-oxides.
  Recent spectroscopic observations of 55 Cancri-e by \cite{tsiaras2015detection} suggest an
  atmosphere which has C/O $\sim$ 1 having a high concentration
  of HCN.}

{Our mineral cloud formation models show a large variation in
  particle sizes and compositions for different planetary scenarios and different sets of element abundances as tabulated in Table \ref{dustvolfractions}. The particle sizes
  are found to be the largest in the case of giant gas planet which is
  indicative of the denser atmosphere resulting in more condensing
  material availability. The particle sizes follow the order of Gas
  Giant $>$ HD149\,026b $>$ 55 Cnc e, with the largest particles found
  at the cloud base in each of the cases below which the particles evaporate. The biggest particle sizes are 54 $\mu$m for the solar case, 3.6$\mu$m 
  for the BSE case, and 3 $\mu$m for the BC Giant gas planet case. These values change for 55 Cnc e to 0.5$\mu$m, 0.47$\mu$m and 0.32 $\mu$m for the solar, BSE and BC cases, respectively. For  HD149\,026b, we find maximum mean cloud particle radii of 0.6$\mu$m, 0.4$\mu$m, and 0.28$\mu$m solar, BSE and BC cases, respectively.  
  The cloud particle number densities
  (n$_{\rm d}$ [cm$^{-3}$}) are  10 times higher for the atmosphere of 55
  Cnc e as compared to HD149\,026b which will contribute to a higher
  cloud opacity on 55 Cnc e. 
}

{We observe a similarity in the cloud property trend
  due to the changes in element abundances on each of the planets
  cases where cloud formation happens. The particle sizes are found to
  be maximum with a solar and BSE composition for each of the
  planets. This is due to higher Mg and O content in the atmosphere
  with these abundances which results in faster material growth  of
  species such as MgSiO$_3$[s] and Mg$_2$SiO$_4$[s]. The dust number
  density (n$_{\rm d}$) is found to be the highest for a MORB atmosphere which
  can be explained due to higher Ti content which results in more seed
  species (TiO$_2$) formation during the nucleation stage. The
  particle composition also follows a similar trend due to changes in
  element abundances. Mineral cloud particles in BC, UC and MORB atmospheres
  predominantly constitute of SiO[s] and SiO$_2$[s] species and their
  percentages vary proportionally with the changes in element
  abundances. Similarly a BSE and solar atmosphere would have dust
  particles predominantly of Mg$_2$SiO$_4$[s], MgSiO$_3$[s] and
  MgO[s]. Only the atmosphere of gas giant has pressures and
  temperatures suitable for cloud base to form particles of
  Al$_2$O$_3$[s] and Fe[s] species. \cite{lee2015modelling} in their
  analysis for 3D cloud formations on HD189\,733b find similar high
  volume fractions of Fe- and Al-binding in their cloud particles and suggest a locally
  lower cloud opacity due to Al$_2$O$_3$[s] and high opacity due to Fe[s]
   which could alter the radiation propagation.}

\noindent
To summarize the results obtained in this work:\\
- Our models suggest possibility of mineral cloud formations on 55 Cnc e and HD149\,026b. The atmosphere of CoRoT-7b or HRSE is found to be too warm on the day-side  for gas condensation to occur based on the atmosphere profile applied.\\  
- Our results indicate that changes in element abundance compositions result in significant changes in the compositions of (mineral) cloud particles. Solar and BSE atmospheres consists primarily of Mg materials whereas BC, UC and MORB atmospheres consist of Si- and Fe-binding materials \\
- The cloud particle properties for different compositions follow a trend of variation independent of local gas-phase pressures and temperatures. As an example, an atmosphere with high abundance of a seed forming species, such as Ti that is found in MORB composition, will have a larger cloud number density as compared to BSE, BC and UC compositions due to higher number of seed particles formed.\\
- The atmospheres of 55 Cnc e and HD149\,026b requires strong element replenishment to be able to sustain cloud formation  in the atmosphere. An atmosphere with no vertical replenishment results in the heavy elements raining out, rendering the atmosphere depleted of heavy elements. 

Magma (or volcano) planets will form thick opaque clouds that will effect the evolution of these rocky planets, and with that maybe also the emergence of plate tectonics. With the  bulk element abundances for extrasolar planets being very difficult to determine from present formation and evolutionary models, Table \ref{dustvolfractions}  provides global mean values for cloud properties for various rocky abundances as a first guidance for retrieval methods  and for planet evolution model. 

\section*{Acknowledgements}

{We thank the referee for providing helpful feedback which led to further improvements of this paper. GM acknowledges an ERASMUS studentship from the TU Delft. ChH
  highlight financial support of the European Community under the FP7
  by an ERC starting grant number 257431. YM greatly appreciates the
  CNES travel funding and post-doctoral fellowship program. We are grateful to J. Fortney for finding us the HD149026b atmosphere structures. We would like to thank Paul Rimmer for useful discussions that helped to improve the manuscript.}




\bibliographystyle{mnras}
\bibliography{bib}





\appendix

\section{Element abundances and growth surface reactions}

\subsection{Terrestrial element abundances from rocks}\label{ss:appA}

The composition of a rock is expressed in terms of weight oxides in
the geoscience literature. This is expressed in weight percentage of a
particular compound, for example the upper crust of the Earth is a
mixture of 66.6$\%$ SiO$_{2}$[s], 15.4$\%$ Al$_{2}$O$_{3}$[s],
3.59$\%$ CaO[s].
In astronomical literature, element
abundances are expressed relative to $10^{12}$ H atoms such the the
element abundance $\epsilon_{\rm i}$ for a certain species $i$ is (i =
Si, O, Mg, $\ldots$)
\begin{equation}
    \mathrm{log (\epsilon_{\rm i}) = log(\frac{n_{\rm i}}{n_{\rm H}}) + 12.}
    \label{eq:ea}
\end{equation}
$n_{\rm i}$ [cm$^{-3}$] is the number density of the element $i$,
hence, n$_{\rm H}$ = 10$^{12}$.  The weight oxide($\%$) of each material
(SiO$_{2}$[s], FeO[s], Al$_{2}$O$_{3}$[s] etc.) is used to calculate
the number of moles of that compound in a particular rock type as
\begin{equation}
    \mathrm{M(No. of moles)} = \frac{\mathrm{m(\%\, of\, Compound)}}{\mathrm{n(molar\,mass [g/mole])}}.
    \label{eq:moles}
\end{equation}
The number of moles of a particular element, e.g. oxygen(O) in SiO$_{2}$[s] is calculated as
\begin{equation}
    \mathrm{M_{O} = \nu_{O} \cdot M}.
    \label{eq:sto}
\end{equation}
 $\nu_{\rm i}$ is the stoichiometric coefficient of the chosen element in the compound. The total number of atoms of a particular element can then be derived from
\begin{equation}
    \mathrm{N_{tot,i} = N_{A} \sum \nu_{s} M_{i, s}.}
    \label{eq:tot}
\end{equation}
n$_{\rm tot,i}$ is the total number of moles of a particular element
in the mixtures, $s$ is the compound (SiO$_{2}$[s],
Al$_{2}$O$_{3}$[s], FeO[s] etc.) and N$_{A}$[atoms/mole] is
Avogadro's number.

The rock composition in Tables~\ref{tab:EAa},~\ref{tab:EAb},  do
not contain information about potential hydrogen content that a rocky
planet's atmosphere may have.
Thus, we follow the method generally used to calculate the element
abundances from meteorites where the abundance ratios are calculated with
respect to the Si atoms in a scale of log$_{10} \epsilon_{\rm Si} =
6$,
\begin{equation}
    \mathrm{log (\epsilon_{\rm Si}) = log(\frac{n_{\rm tot,i}}{n_{\rm Si}}) + 6.}    
\label{eq:si}
\end{equation}

To be able to convert the Si-normalized abundances to a H-normalized
abundances scale, we follow the method outlined by
\cite{Palme2013}. The conversion factor between the two scales was
calculated by dividing the H-normalized solar abundances by the
Si-normalized meteorite abundances. The comparison was made for all
elements with an error of the corresponding photospheric abundance of
less than 0.1 dex, i.e., less than 25$\%$. The log of the average
ratio of solar abundance per $10^{12}$ H atoms/meteorite abundance per
$10^{6}$ silicon atoms is 1.546 $\pm$ 0.045 \citep{Palme2013}. Thus,
\begin{equation}
    \mathrm{log(\epsilon_{H})} = \mathrm{log(\epsilon_{Si})} + 1.546
    \label{eq:conv}
\end{equation}

\begin{figure}
\hspace*{-0.5cm}
\centering
 \includegraphics[scale=0.41]{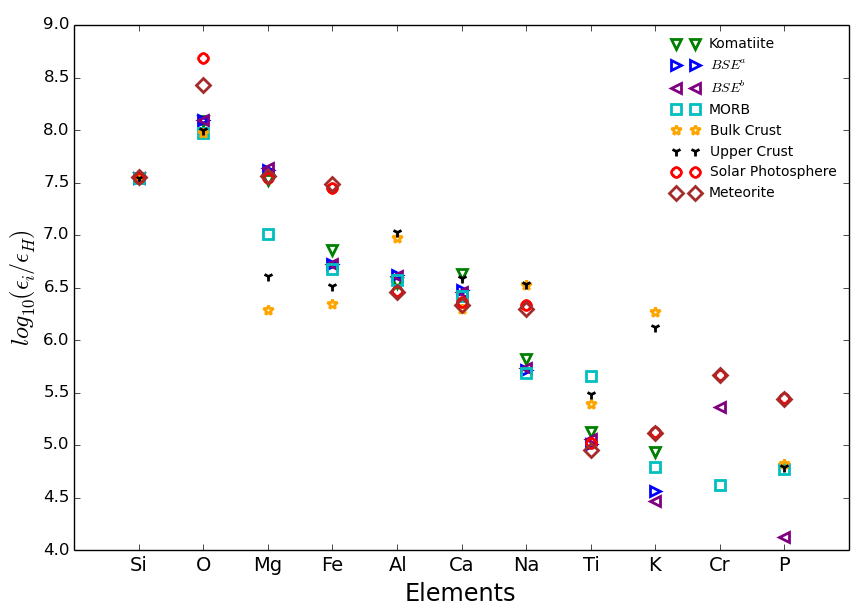}
 \caption{Element abundance (log$_{10}(\epsilon_{\rm H})$ = 12 scale),
   for six Earth rock and magma composition. The komatiite and
   BSE$^{a}$ compositions are used by \citealt{miguel2011} and
   BSE$^{b}$, MORB, Bulk Crust and Upper Crust compositions are used
   by \citealt{Itoetal} (see Tables~\ref{tab:EAa},~\ref{tab:EAb}). The
   solar photospheric and meteorites abundances are shown for
   comparison (\citealt{grevesse2007solar}).}
 \label{fig:EA}
\end{figure}

The resulting element abundances in the log($\epsilon_{H}$) = 12 scale
are listed in Table~\ref{tab:EAb}. In Fig.~\ref{fig:EA}, we compare the
element abundances derived for all the six types of rocks (bulk crust
-- yellow stars, upper crust -- black, MORB -- light-blue squares
(metal-oxide-ridge basalt), BSE$^{\rm a}$ -- blue triangle (bulk
silicate Earth), BSE$^{\rm b}$ -- purple triangle, komatite -- green
triangle). We include a comparison to the solar (red circles) and the
meteorite (brown diamonds) element abundances. 

\begin{table*}
\centering
\caption{Weight ($\%$) of oxides for various types of Earth and magma rocks.}
\resizebox{0.8\linewidth}{!}{%
\begin{tabular}{c c c c c c c}

    \hline \hline
    Weight oxide ($\%$) & Komatite$^{a}$ 
    & BSE$^{b}$ & BSE$^{c}$ & MORB$^{c}$ 
    & Upper Crust$^{d}$  & Bulk Crust$^{d}$\\
    \hline

    SiO$_{2}$ & 47.10 & 45.97 & 45.1 & 49.6 & 66.6 & 60.6\\
    MgO & 29.60 & 36.66 & 37.9 & 9.75 & 2.5 & 4.\\
    FeO & - & 8.24 & 8.06 & 8.06 & 5.04 & 6.71\\
    Al$_{2}$O$_{3}$ & 4.04 & 4.77 & 4.46 & 16.8 & 15.4 & 15.9\\
    CaO & 5.44 & 3.78 & 3.55 & 12.5 & 3.59 & 6.40\\
    Na$_{2}$O & 0.46 & 0.35 & 0.36 & 2.18 & 3.27 & 3.07\\
    Cr$_{2}$O$_{3}$ & - & - & 0.38 & 0.07 & 0.0003 & 0.0004\\
    TiO$_{2}$ & 0.24 & 0.18 & 0.2 & 0.9 & 0.64 & 0.72\\
    K$_{2}$O & 0.09 & 0.04 & 0.03 & 0.07 & 2.8 & 1.81\\
    P$_{2}$O$_{5}$ & - & - & 0.02 & 0.1 & 0.13 & 0.15\\
    Fe$_{2}$O$_{3}$ & 12.8 & - & - & -  & - & -\\
    
    \hline
    
\end{tabular}}
{\ }\\
{\small
$^{a}$ \cite{2004Icar..169..216S},
$^{b}$ O'Neill $\&$ Palme (1998),
$^{c}$ \cite{McDonough1995223},
$^{d}$ \cite{2009pctc.book.....T}.
$^{e}$ \cite{grevesse2007solar}.}

 \label{tab:EAa}
\end{table*}

\subsection{Surface-growth reactions}\label{ss:Ap}
\begin{table*}
\caption{Extended set of surface growth reactions for the growth of dust particles in addition to Table~1 in Helling et al. (2008b). The total sum of surface growth reaction taken into account here is 79.} 
\begin{tabular}{c c l l}

    \hline \hline
   Index $r$  &  Solid $s$ & Surface reaction & Key species\\
    \hline

    61. &  MgO[s] & 2MgH + 2H$_2$O $\longrightarrow$  2MgO[s] + 3H$_2$ & MgH \\
    62. & MgSiO$_{3}$[s] & 2MgH + 2SiO + 4H$_2$O $\longrightarrow$ 2MgSiO$_3$[s] + 5H$_2$ &  min\{MgH, SiO\}\\
    63. & &  MgH + SiH + 3H$_2$O $\longrightarrow$ MgSiO$_3$[s] + 4H$_2$ & min\{MgH, SiH\} \\
    64. & & 2MgH + 2SiN + 6H$_2$O $\longrightarrow$ 2MgSiO$_3$[s] + 7H$_2$ + N$_2$ &  min\{MgH, SiN\} \\
    65. & & MgS + Si + 3H$_2$O $\longrightarrow$ MgSiO$_3$[s] + H$_2$S + 2H$_2$ &  min\{MgS, Si\}\\
    66. & & 2MgN + 2Si + 3H$_2$O $\longrightarrow$ 2MgSiO$_3$[s] + 3H$_2$ + N$_2$ &  min\{MgN, Si\}\\
    67. & Mg$_2$SiO$_4$[s] & 2MgH + SiO + 3H$_2$O $\longrightarrow$ Mg$_2$SiO$_4$[s] + 4H$_2$ &  min\{2MgH, SiO\}\\
    68. & & 4MgH + 2SiH + 8H$_2$O $\longrightarrow$ 2Mg$_2$SiO$_4$[s] + 11H$_2$ &  min\{2MgH, SiH\}\\
    69. & & 4MgH + 2SiN + 8H$_2$O $\longrightarrow$ 2Mg$_2$SiO$_4$[s] + N$_2$ + 10H$_2$ & min\{2MgH, SiN\} \\
    70. & & 2MgS + Si + 4H$_2$O $\longrightarrow$ Mg$_2$SiO$_4$[s] + 2H$_2$S + 2H$_2$ &  min\{2MgS, Si\}\\
    71. & & 2MgN + Si + 4H$_2$O $\longrightarrow$ 2Mg$_2$SiO$_4$[s] + N$_2$ + 4H$_2$ &  min\{2MgN, Si\}\\
    72. & SiO$_2$[s] & 2SiH + 4H$_2$O $\longrightarrow$ 2SiO$_2$[s] + 5H$_2$ & SiO\\
    73. & & 2SiN + 4H$_2$O $\longrightarrow$ 2SiO$_2$[s] + N2 + 4H$_2$ &  SiN\\
    74. & SiO[s] & 2SiH + 2H$_2$O $\longrightarrow$ 2SiO[s] + 3H$_2$ & SiO\\
    75. & & 2SiN + 2H$_2$O $\longrightarrow$ 2SiO[s] + N$_2$ + 2H$_2$ & SiN\\
    76. & Fe[s] & 2FeH + H$_2$ $\longrightarrow$ 2Fe[s] + 2H$_2$ & FeH\\
    77. & FeO[s] & 2FeH + 2H$_2$O $\longrightarrow$ 2FeO[s] + 3H$_2$ &  FeH\\
    78. & FeS[s]& 2FeH + 2H$_2$O $\longrightarrow$ 2FeS[s] + 3H$_2$ & FeH\\
    79. & Fe$_2$O$_3$[s] & 2FeH + 3H$_2$O $\longrightarrow$ Fe$_2$O$_3$[s] + 4H$_2$ & FeH\\
    \hline
\end{tabular} 
\label{table:reactions}
\end{table*}



\bsp	
\label{lastpage}
\end{document}